\documentclass[journal]{IEEEtran}
\PassOptionsToPackage{ruled}{algorithm2e}
\usepackage[T1]{fontenc}
\usepackage[utf8]{inputenc}
\usepackage{algorithm2e}
\usepackage{amsmath}
\usepackage{amsthm}
\usepackage{amssymb}
\usepackage{comment}
\usepackage{xcolor}
\allowdisplaybreaks
\usepackage{graphicx}
\usepackage[bookmarks=true,bookmarksnumbered=true,bookmarksopen=true,bookmarksopenlevel=1,
 breaklinks=false,pdfborder={0 0 0},pdfborderstyle={},backref=false,colorlinks=false]
 {hyperref}
\hypersetup{pdftitle={Your Title},
 pdfauthor={Your Name},
 pdfpagelayout=OneColumn, pdfnewwindow=true, pdfstartview=XYZ, plainpages=false}
\theoremstyle{plain}
\newtheorem{thm}{\protect\theoremname}
\theoremstyle{plain}
\newtheorem{lem}[thm]{\protect\lemmaname}
\usepackage[caption=false,font=small]{subfig}
\usepackage{cite}
\usepackage{balance}

\usepackage{cite}
\usepackage{acronym}
\AtBeginDocument{\acrodef{6G}{sixth-generation}
\acrodef{ALM}{augmented Lagrangian method}
\acrodef{AO}{alternating optimization}
\acrodef{AoA}{angle of arrival}
\acrodef{AoD}{angle of departure}
\acrodef{APGM}{alternating projected gradient method}
\acrodef{APM}{accelerated proximal gradient method}
\acrodef{AP}{access point}
\acrodef{ASP}{antenna separation product}
\acrodef{AWGN}{additive white Gaussian noise}
\acrodef{BC}{broadcast channel}
\acrodef{BCM}{block coordinate maximization}
\acrodef{BEP}{bit error probability}
\acrodef{BER}{bit error rate}
\acrodef{BF-MIMO}[BF\mbox{-}MIMO]{beamforming MIMO}
\acrodef{BF}{beamforming}
\acrodef{BS}{base station}
\acrodef{bpcu}{bits per channel use}
\acrodef{CFO}{carrier frequency offset}
\acrodef{CP}{cyclic prefix}
\acrodef{CPU}{central processing unit}
\acrodef{CR}{communication rate}
\acrodef{CSI}{channel state information}
\acrodef{CSIR}{channel state information at RX}
\acrodef{SSK}{space shift keying}
\acrodef{CRLB}{Cram\'er-Rao lower bound}
\acrodef{CSIT}{channel state information at TX}
\acrodef{DCMC}{discrete\mbox{-}input continuous\mbox{-}output memoryless channel}
\acrodef{DFT}{discrete Fourier transform}
\acrodef{DL}{deep learning}
\acrodef{DL-TR-GSM}{dual-layered transmit-receive \acl{GSM}}
\acrodef{DLT}{dual-layered transmission}
\acrodef{DMA}{dynamic metasurface antenna}
\acrodef{DOA}{direction of arrival}
\acrodef{DoF}{degrees of freedom}
\acrodef{DNN}{deep neural network}
\acrodef{DPC}{dirty paper coding}
\acrodef{DRL}{deep reinforcement learning}
\acrodef{EE}{energy efficiency}
\acrodef{EGC}{equal gain combining}
\acrodef{EM}{electromagnetic}
\acrodef{EVD}{eigenvalue decomposition}
\acrodef{FPGA}{field programmable gate array}
\acrodef{FSPL}{free space path loss}
\acrodef{FFT}{fast Fourier transform}
\acrodef{FDE}{frequency domain equalization}
\acrodef{GRSM}{generalized \acl{RSM}}
\acrodef{GSM}{generalized \acl{SM}}
\acrodef{HMIMO}{holographic MIMO}
\acrodef{IA}{inner approximation}
\acrodef{IFFT}{invserse fast Fourier transform}
\acrodef{ICI}{inter-channel interference}
\acrodef{iid}[i.i.d.]{independent and identically distributed}
\acrodef{IMT}{International Mobile Telecommunications}
\acrodef{IQ}{in\mbox{-}phase and quadrature}
\acrodef{ISAC}{integrated sensing and communication}
\acrodef{ISI}{intersymbol interference}
\acrodef{ISI-free}[ISI\mbox{-}free]{intersymbol interference free}
\acrodef{LIS}{large intelligent surface}
\acrodef{LOS}{line\mbox{-}of\mbox{-}sight}
\acrodef{KKT}{Karush\mbox{-}Kuhn\mbox{-}Tucker} 
\acrodef{MA}{movable antenna}
\acrodef{MAC}{multiple-access channel}
\acrodef{mmWave}{millimeter-wave}
\acrodef{MI}{mutual information}
\acrodef{MIMO}{multiple\mbox{-}input multiple\mbox{-}output}
\acrodef{mMIMO}{massive MIMO}
\acrodef{MISO}{multiple\mbox{-}input single\mbox{-}output}
\acrodef{ML}{maximum likelihood}
\acrodef{MRC}{maximal ratio combining}
\acrodef{MMSE}{minimum mean square error}
\acrodef{MU-TR-GSM}{multiuser transmit-receive  \acl{GSM} }
\acrodef{NCSIT}{no channel state information at TX}
\acrodef{NLOS}{non\mbox{-}\acs{LOS}} 
\acrodef{NOMA}{non-orthogonal multiple access}
\acrodef{OFDM}{orthogonal frequency division multiplexing}
\acrodef{OFDMA}{orthogonal frequency division multiple access}
\acrodef{OMP}{orthogonal matching pursuit}
\acrodef{OTFS}{orthogonal time frequency space}
\acrodef{UAV}{unmanned aerial vehicle}
\acrodef{umMIMO}{ultra-massive MIMO}
\acrodef{PA}{power amplifier}
\acrodef{PAE}{power added efficiency}
\acrodef{PAPR}{peak\mbox{-}to\mbox{-}average power ratio}
\acrodef{PDF}{probability density function}
\acrodef{PEP}{pairwise error probability}
\acrodefplural{PEP}{pairwise error probabilities}
\acrodef{PGM}{projected gradient method}
\acrodef{PMP}{probability mass function}
\acrodef{PSM}{precoding-aided spatial modulation}
\acrodef{QSM}{quadrature spatial modulation}
\acrodef{RC}{reorganization computation}
\acrodef{RCS}{radar cross section}
\acrodef{RF}{radio frequency}
\acrodef{RHS}{right-hand side}
\acrodef{RIS}{reconfigurable intelligent surface}
\acrodef{RSM}{receive spatial modulation}
\acrodef{RX}{receiver}
\acrodef{SDR}{semi-definite relaxation}
\acrodef{SE}{spectral efficiency}
\acrodef{SEP}{symbol error probability}
\acrodef{SER}{symbol error rate}
\acrodef{SIC}{successive interference cancellation}
\acrodef{SIM}{stacked intelligent metasurface}
\acrodef{SINR}{signal-to-interference-plus-noise ratio}
\acrodef{SISO}{single-input single-output}
\acrodef{SM}{spatial modulation}
\acrodef{SMX-MIMO}[SMX\mbox{-}MIMO]{spatial multiplexing MIMO}
\acrodef{SMX}{spatial multiplexing}
\acrodef{SNR}{signal-to-noise ratio}
\acrodef{SC}{single carrier}
\acrodef{SCA}{successive convex approximation}
\acrodef{SVD}{singular value decomposition}
\acrodef{SPST}{single pole single-throw}
\acrodef{SR}{sensing rate}
\acrodef{SU}{secondary user}
\acrodef{TDE}{time domain equalization}
\acrodef{THz}{terraherz}
\acrodef{TX}{transmitter}
\acrodef{ULA}{uniform linear array}
\acrodef{URA}{uniform rectangular array}
\acrodef{VGA}{variable gain amplifier}
\acrodef{WSR}{weighted sum rate}
\acrodef{wrt}[w.r.t.]{with respect to}
\acrodef{ZF}{zero-forcing}
\acrodef{ZMCG}{zero-mean complex Gaussian}

}
\usepackage[skip=0pt]{caption} 
\makeatletter 
\pretocmd\@bibitem{\color{black}\csname keycolor#1\endcsname}{}{\fail}
\newcommand\citecolor[1]{\@namedef{keycolor#1}{ \color{blue}}}
\makeatother
\providecommand{\lemmaname}{Lemma}
\providecommand{\theoremname}{Theorem}

\begin{document}
\title{Beamforming Design for Multi-user MIMO Near-Field ISAC Systems with Movable Antennas}
\author{Nemanja Stefan Perovi\'c,~\IEEEmembership{Member,~IEEE}, Keshav Singh,~\IEEEmembership{Senior Member, IEEE}, \\Chih-Peng Li,~\IEEEmembership{Fellow, IEEE},   and Mark F. Flanagan,~\IEEEmembership{Senior Member, IEEE}

\thanks{This work was supported in part by the National Science and Technology Council of Taiwan under Grants NSTC 115-2221-E-110-036-MY3, NSTC 115-2640-E-110-003, and NSTC 115-2221-E-110-035-MY2, and also in part by the Sixth Generation Communication and Sensing Research Center funded by the Higher Education SPROUT Project, the Ministry of Education of Taiwan. (\textit{Corresponding author: Keshav Singh).}}

\thanks{N. S. Perovi\'c, K. Singh, and C.-P. Li are with the Institute of Communications Engineering, National Sun Yat Sen University, Kaohsiung 80424, Taiwan (e-mail:  \{n.s.perovic,\,keshav.singh,\,cpli\}@mail.nsysu.edu.tw).} 

\thanks{M. F. Flanagan is with the School of Electrical and Electronic Engineering, University College Dublin, Dublin 4, D04 V1W8, Ireland (e-mail: mark.flanagan@ieee.org).}

\thanks{A partial version of this work was presented at the IEEE International Conference on Communications 2026, Glasgow, Scotland, UK \cite{perovic2025weighted}.}

}
\maketitle
\begin{abstract}
\Ac{ISAC} has been recognized as one of the key technologies capable of simultaneously improving communication and sensing services in future wireless networks. Moreover, the introduction of recently developed \acp{MA} has the potential to further increase the performance gains of ISAC systems. Although the gains of MA-enabled ISAC systems are relatively well studied in the far field, they remain almost unexplored in near-field scenarios. Motivated by this, in this paper we maximize the \ac{WSR} for communication users while maintaining a minimum sensing requirement in an MA-enabled near-field ISAC system. To achieve this goal, we propose algorithms that optimize the communication precoding matrices, the sensing transmit beamformer, the sensing receive combiner, the positions of the users’ MAs and the positions of the \ac{BS} transmit MAs in an alternating manner for the considered ISAC system, for the cases where linear procoding and \ac{ZF} precoding are employed at the \ac{BS}. Simulation results show that using MAs in near-field ISAC systems provides a substantial performance advantage compared to near-field ISAC systems equipped with fixed antennas only. We show that the  scheme with linear precoding achieves larger WSR for unequal users' weight rates, while the scheme with ZF precoding maintains an approximately constant WSR for all users' weight rates. Additionally, we demonstrate that the WSRs of the proposed schemes are highly dependent on the inter-antenna interference between different user’s MAs, and that the sensing performance is significantly more affected by the minimum sensing \ac{SINR} threshold compared to the communication performance. 
\acresetall{}
\end{abstract}

\IEEEpeerreviewmaketitle{}

\section{Introduction}
\bstctlcite{BSTcontrol}
\IEEEPARstart{F}{uture} wireless communication systems will have to support many new
functionalities, among which high-precision sensing is one of the
most important as it enables various environment-aware applications
such as augmented reality and digital twins. A promising technology
for implementing this functionality is that of \ac{ISAC} \cite{liu2022integrated};
this refers to a design paradigm in which sensing and communication
systems are integrated to efficiently utilize the shared spectrum
and hardware resources, while offering mutual benefits \cite{perovic2025sensing}.
Through collaboration that includes appropriate reuse of resources
and information, ISAC enhances both sensing and communication capabilities,
and provides flexible trade-offs between these two functionalities
in different use cases \cite{magbool2025survey}. Furthermore, 
due to the possibility of integrating new wireless technologies, such as radio-frequency identification (RFID) \cite{rojith2025optimizing}, 
\acp{UAV} \cite{saikia2024hybrid}, fluid antennas \cite{11184593}, and \ac{NOMA} \cite{mondal2025outage}, ISAC has attracted significant attention from both the academic community and industry~\cite{11390054}.

In conventional systems, fixed-position antenna arrays are generally deployed in order to carry out communication and sensing tasks. In particular, \ac{MIMO} technology is a promising
solution for ISAC due to its ability to steer the communication
and the sensing signals toward the desired directions, reducing the signal interference and thereby improving the overall system
performance \cite{liu2020joint}. As a result, \ac{MIMO} systems can
offer enhanced sensing accuracy and/or channel capacity. Motivated by
this, a significant research effort has been dedicated to exploring the problem of optimal antenna positioning in MIMO systems. An efficient algorithm for selecting
a subset of transmit antennas that maximizes the channel capacity
in a \ac{mMIMO} array was presented in \cite{ibrahim2020fast}. Another technique, known as sparse array synthesis, uses a flexible array structure to achieve different radiation
pattern properties (e.g., desired pattern directivity, reduced power
leakage) \cite{yang2022low,liu2019synthesizing}.
However, the basic assumption underlying both of the aforementioned techniques
is that the set of available antenna positions is fixed and discrete,
which prevents the utilization of the full range of positions of the antennas in the array.

The recent development of metamaterials-enabled manufacturing
of intelligent metasurfaces with densely packed tunable elements,
which can dynamically control the propagation of radio waves, have potential to improve ISAC
capabilities \cite{chepuri2023integrated}. An element-wise closed-form
optimization method for adjusting the \ac{RIS} reflection coefficients
for improving the sensing \ac{SINR} and reducing the communication
interference was proposed in \cite{zhong2023joint}. 
In \cite{yang2024ris},
it was demonstrated that deploying an RIS can significantly improve
the performance of cooperative multicell ISAC systems in reducing
the transmit power, while achieving 
communication and
sensing requirements. Optimization of an ISAC system equipped with
an active RIS was presented in \cite{zhu2023joint}, where it was demonstrated that the use of an active RIS can provide significantly better performance compared to an ISAC system
with a passive RIS. However, all of the designs mentioned above
are based on fixed metasurface element positions and/or orientations
once manufactured. In practical wireless systems with varying signal
propagation environments, such fixed-position metasurfaces  may not fully exploit
the spatial variation of wireless channels, which  limits their
\acp{DoF} in the spatial domain.

Recently, \acp{MA} have emerged as a promising technology to further
improve the effectiveness of ISAC by enabling the local movement of
antennas at transmitters and receivers \cite{zhu2023movable}. Different from conventional
fixed antennas, MAs can adjust their placements to adapt to evolving channel conditions. Dynamic positioning of such
antennas enables precise beamforming design, avoiding undesirable
side lobes and reducing interference, which enhances data rate/reliability
and sensing accuracy \cite{zhu2025tutorial}. Therefore, superior
performance can be achieved by MA systems using the same or even a smaller
number of antennas compared to conventional fixed antenna~arrays.

Motivated by this, a significant number of papers have studied the
use of \acp{MA} in ISAC systems. In \cite{ma2024movable}, the authors
studied the \ac{CRLB} for \ac{AoA} estimation as a function of the
MA positions in 1D and 2D antenna arrays, and proposed algorithms
for its optimization. 
Maximization of the weighted sum
of the communication rate and the sensing \ac{MI} in a bistatic ISAC
system with an MA transmit array was considered
in \cite{lyu2025movable}. An optimization framework that maximizes
the sensing SINR in a multi-user bistatic ISAC system, where MAs are
deployed for adjusting the array responses at both the transmitter
and the receiver of a \ac{BS}, was introduced in \cite{jiang2025movable}. {A two-timescale optimization framework that allows slow MA movements once in a longer scanning period, to reduce the delay impact introduced by the associated electro-mechanical repositioning process, was proposed in \cite{khalili2025movable}.}
In \cite{yang2025robust}, the authors proposed the design and optimization
of an \ac{RIS}-aided ISAC broadcast system with MAs at the \ac{BS}
under imperfect \ac{CSI} for both sensing and communication channels.
Practical gains achieved by using MAs with a discrete set of
possible antenna positions in an ISAC system with a dynamic radar
cross-section were analyzed in \cite{khalili2024advanced}. 
{
In \cite{xiu2025movable}, the authors considered using MAs to improve the robustness of cooperative ISAC networks to imperfect \ac{CSI} and time synchronization errors.
A novel MA-aided cell-free ISAC architecture, capable of enhancing communication rates and reducing the impact of time synchronization errors, was proposed and optimized in \cite{xiu2026robust}.
In \cite{xiu2026meta}, the authors investigated the application of MAs to a similar ISAC system to mitigate impairments caused by \ac{CFO}.}

{
As previously outlined, the use of MAs improves both the communication and the sensing functionality of ISAC. Moreover, these gains generally increase with the size of the MA regions. For example, larger MA regions tend to provide a higher number of \ac{DoF} in ISAC systems. In addition, the use of high-frequency bands offers a twofold benefit. First, the abundant spectrum available in these bands enables communication systems to achieve high data rates. Second, the allocation of wider bandwidths for target sensing enables high-resolution imaging. Hence, combining larger MA regions and high-frequency bands has the \mbox{potential to} provide significant performance gains in MA-enabled ISAC \mbox{systems}. Nevertheless, this combination also expands the near-field region, thereby increasing the importance of near-field effects and requiring new channel modeling and beamforming designs \cite{ma2025movable,liu2023near,an2024near}.
}

To the best of the authors' knowledge, there are only a few papers
dealing with the design of such systems. In \cite{sun2025rotatable},
the authors proposed a sensing-centric and a communication-centric
design for near-field ISAC systems with rotatable MAs at the BS, and
demonstrated their performance advantage compared to fixed-position
antennas and non-rotatable MAs;  
{the presented designs were only applied to users equipped with one fixed antenna.}
In \cite{wang2025movable}, the authors have derived the worst-case \acp{CRLB} for target sensing using 1D and 2D MA arrays in the near field, and developed algorithms for their optimization, {but without any attempt to include the communication functionality in the presented analysis.}
An ISAC framework for near-field MA-aided communication systems, which can reconstruct complete \ac{CSI} and locate scatterers effectively, was proposed in \cite{sun2026movable}.  
{However, it remains unclear if this framework can be implemented in ISAC systems where users' devices are equipped with MAs instead of fixed antennas.} 
The maximization of the weighted sum
of the communication and sensing rates in a full-duplex near-field
ISAC communication system with MAs at the BS was studied in \cite{ding2025movable}.
For the latter system, the MA positions were obtained by selecting
from a predefined set of MA positions those that provide the largest
value of the objective function. 
{The use of predefined MA positions instead of optimized ones implies that a certain performance degradation almost certainly occurs. Also, the paper does not take into account the possibility that users may utilize MAs.}
A deep learning (DL)-based optimization framework was proposed in \cite{zhang2025deep} to maximize the \ac{WSR} while guaranteeing the required sensing performance through \ac{CRLB} constraints for an \ac{RIS} architecture consisting~of multiple stacked movable elements operating in the near-field~regime.

Against this background, the contributions of this paper are summarized
as follows:
\begin{enumerate}
\item We introduce an optimization framework for multi-user
ISAC systems operating in the near field, where the BS transmitter and each user are equipped with multiple MAs. Within this framework,
we formulate an optimization problem with the aim of
maximizing the WSR while satisfying minimum sensing SINR requirements\footnote{The sensing SINR determines the target detection probability, as well as the sensing \ac{MI} (i.e., the "amount of information" obtained through the sensing process). Moreover, using the closed-form expression for the sensing SINR makes the optimization of the considered ISAC system more analytically tractable. Due to these reasons, we have adopted the sensing SINR as the sensing performance metric.}, for the cases of both linear and \ac{ZF} precoding\footnote{In the partial version of this paper in  \cite{perovic2025weighted}, only the optimization of a near-field ISAC system with linear precoding was studied. Also, in  \cite{perovic2025weighted} only communication users were equipped with MAs, while in this paper both the BS and users have MAs.} for the communication signals at the BS.

\item For the system with linear precoding, we propose an \ac{AO} based algorithm to solve the formulated problem by decomposing it into multiple subproblems\footnote{{We assume that target tracking is implemented and that the previously estimated target position is used as the current target position by the proposed algorithm.}}. To optimize the sensing receive combiner, we provide a closed-form solution. The communication precoding matrices are optimized
using the \ac{SCA} method, which provides a tight concave lower bound on the achievable rate of each user. Using the same approach and employing \ac{SDR}, we optimize the sensing transmit beamforming vector. Finally, the positions of the users' MAs are obtained via the \ac{PGM}, while the positions of the BS transmit MAs are optimized by the \ac{ALM}. A similar \ac{AO} approach is proposed to solve the 
problem for the system with \ac{ZF} precoding. However, in this case the positions of the users' MAs are optimized by the \ac{ALM}, since they also affect the sensing SINR.
\item We show through simulations that the proposed algorithms achieve significantly improved performance compared to benchmark schemes equipped with only fixed antennas, due to a larger number of \ac{DoF}. Moreover,
we demonstrate that a larger WSR for the scheme with linear precoding is obtained when the users' rate weights are unbalanced, while the WSR for the scheme with ZF precoding
stays approximately constant irrespective of these rate weights. In systems with a low number of MAs per user, ZF precoding can provide the best WSR performance, while on the other hand this performance decreases significantly for a larger number of MAs per user due to larger inter-antenna interference between different user’s MAs. {In addition, increasing the sizes of the BS and user MA regions can significantly enhance the performance of schemes with ZF precoding relative to schemes with linear precoding.}
Finally, we demonstrate that the sensing performance for both precoding schemes in the considered ISAC system is significantly more affected by the sensing SINR threshold
compared to the communication performance.
\end{enumerate}

The rest of this paper is organized as follows. In Section II, we introduce the system model and formulate the optimization problem to maximize the WSR of an ISAC MIMO system with MAs at the BS transmitter and at each user. In Section III, we propose and derive an algorithm to solve this optimization problem in the case of linear precoding. An algorithm to solve the same optimization problem in the case of ZF precoding is proposed in Section IV. In Section V, we present numerical WSR results for the proposed algorithms, and use these to illustrate their relative advantages. Finally, Section VI concludes this paper.

\textit{Notation}: Bold lower and upper case letters represent vectors and matrices, respectively. $\mathbf{I}_{x}$ is the identity matrix of size $x\times x$. $\text{Tr}(\mathbf{X})$, $\text{rank}(\mathbf{X})$, $||\mathbf{X}||$ and $|\mathbf{X}|$ denote the trace, rank, norm and determinant of matrix $\mathbf{X}$, respectively. $\mathbf{X}\succeq(\succ)\mathbf{Y}$ means that $\mathbf{X}-\mathbf{Y}$  is positive semidefinite (definite). $\mathbb{E}\{\cdot\}$ stands for the expectation operator. $\ln(\cdot)$ is the natural logarithm, and $(\cdot)^{T}$ and $(\cdot)^{H}$ represent transpose and Hermitian transpose, respectively. $\mathcal{CN}(\mu,\sigma^{2})$ denotes a circularly symmetric complex Gaussian random variable with mean $\mu$  and variance $\sigma^{2}$. $\mathfrak{R}(\cdot)$ and $\mathfrak{I}(\cdot)$ denote the real and imaginary part of a complex number, respectively. $\mathbf{X}(i,k)$ denotes the \textit{k}-th element of the $\textit{i}$-th row of matrix $\mathbf{X}$. $\mathbf{x}(i)$ denotes the \textit{i}-th element of vector $\textbf{x}$.

\section{System Model}
\begin{figure}[t]
\centering
\includegraphics[width=7cm, height=6cm]{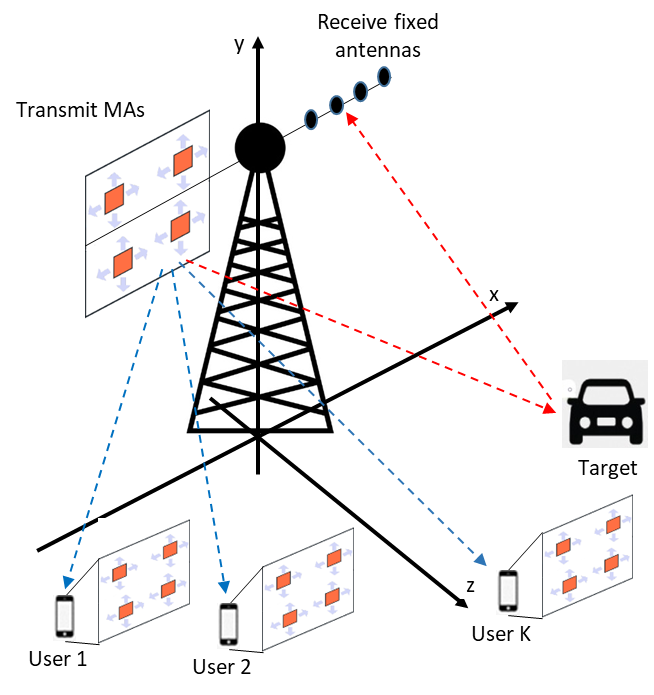}\caption{{An illustration of the MA-aided near-field ISAC system. \label{fig:sys_mod}}}
\end{figure}

In the considered MA-aided near-field ISAC system, a dual-functional
BS is equipped with a transmit MA array, as shown in Fig. \ref{fig:sys_mod}.  This array with $N_{t}$
antennas transmits the ISAC signal, which enables communication with
$K$ users and detection of a sensing target. Each of these MAs can
be moved within a square transmit region which is denoted as $C_{t}$.
In addition, the BS has a second antenna array with $N_{r}$ fixed
antennas, which receive the reflected echo signal from the sensing
target. These two arrays are separated one from another so that the
mutual coupling between antennas from different arrays is negligible.
Each user is equipped with $N_{u}$ receive MAs, which can be moved within the square region denoted as $C_{k}$.

To describe the geometry of the considered system, we establish a 3D
Cartesian coordinate system, where the $xz$-plane represents the
ground. The BS is placed at the origin with both of its antenna arrays
in the $xy$-plane. The BS MA transmit region $C_{t}$ has its center
at $\textbf{o}_{t}=[x_{t},y_{t},0]^{T}$ and its side length is $L_{t}$.
The coordinates of the $m$-th ($1\le m\le N_{t}$) transmit antenna
are given by $\mathbf{t}_{m}=[x_{t,m},y_{t,m},0]^{T}$. The receive
BS antennas are placed in an \ac{ULA} parallel to the \mbox{$x$-axis}, with the
midpoint at $\textbf{o}_{r}=[x_{r},y_{r},0]^{T}$ and a length of $L_{r}$.
The coordinates of the $n$-th ($1\le n\le N_{r}$) receive antenna
are $\mathbf{r}_{n}=[x_{r,n},y_{r},0]^{T}$, where $x_{r,n}=x_{r}-L_{r}/2+(n-1)L_{r}/(N_{r}-1)$.
Without loss of generality, we assume the MA regions of all users
are parallel to the $xy$-plane. For user $k$, the region $C_{k}$
is of a square shape with the center at $\textbf{o}_{u,k}=[x_{u,k},y_{u,k},z_{u,k}]^{T}$
and its side is denoted as $a_{k}$. The coordinates of the $N_{u}$ MAs
in $C_{k}$ are collectively denoted as $\mathbf{q}_{k}=[(\mathbf{q}_{k,1})^{T},(\mathbf{q}_{k,2})^{T},\dots,(\mathbf{q}_{k,N_{u}})^{T}]^{T}$,
where $\mathbf{q}_{k,b}=[x_{k,b},y_{k,b},z_{k}]^{T}\in C_{k}$ for
$1\le b\le N_{u}$ \footnote{{We have to emphasize that the presented optimization framework can be implemented even when the users' MA regions are arbitrary oriented. In this scenario, the position of the $b$-th MA of user $k$ in the global coordinate system can be expressed as $\mathbf{q}_{k,b}=\mathbf{o}_{u,k}+\mathbf{R}_{k}\mathbf{u}_{k,b}$, where $\mathbf{u}_{k,b}$ denotes the position of the same MA in the local coordinate system of that user and the orthogonal matrix $\mathbf{R}_{k}\in\mathbb{R}^{3\times3}$ specifies the orientation of the user’s device \cite{ding2024near}. Since the whole region $C_{k}$ is in one plane, the position of each MA can be determined by two coordinates, while the third coordinate is always zero in the local coordinate system. The main difference is that for determining the optimal positions of the user's MAs, the appropriate gradients have to be calculated w.r.t. the local (instead of the global) coordinates. In our paper when the MA regions of the BS and users are parallel (i.e., $\mathbf{R}_{k}=\mathbf{I}_{3}$ for all users), we have $\mathrm{d}\mathbf{q}_{k,b}=\mathrm{d}\mathbf{u}_{k,b}$, so the previously mentioned gradients can be equivalently calculated w.r.t. the local or the global coordinates of user's MAs (as is done in the paper). Furthermore, due to the near field beam-focusing ability to direct signal energy to a specific spot instead to a specific direction as in far field beamforming, it is expected that the system performance is less affected by the users' orientations in the near field than in the far field.}}. Finally, the position of the sensing target is
specified as $\text{\ensuremath{\mathbf{s}}}=[x_{s},y_{s},z_{s}]^{T}$.

Since the users are located in the near-field region of the BS, the quasi-static
spherical wave channel model is used for the relevant communication channels
\cite{liu2023near}. Accordingly, the channel between the BS and user
$k$ can be expressed as 
\begin{equation}
\mathbf{H}_{k}=\rho_{k}\left[\begin{array}{ccc}
e^{j\frac{2\pi}{\lambda}\|\mathbf{t}_{1}-\mathbf{q}_{k,1}\|} & \cdots & e^{j\frac{2\pi}{\lambda}\|\mathbf{t}_{N_{t}}-\mathbf{q}_{k,1}\|}\\
\vdots & \ddots & \vdots\\
e^{j\frac{2\pi}{\lambda}\|\mathbf{t}_{1}-\mathbf{q}_{k,N_{u}}\|} & \cdots & e^{j\frac{2\pi}{\lambda}\|\mathbf{t}_{N_{t}}-\mathbf{q}_{k,N_{u}}\|}
\end{array}\right],
\end{equation}
where $\lambda$ is the wavelength and $\rho_{k}$ is the free space
path-loss\footnote{The near-field communication channels are modeled by the uniform spherical wave model, in which
the free space path-losses between different transmit BS MAs and MAs of a single user have approximately the same amplitude \cite{liu2023near}.}.

For a point target located in the near-field region of the BS, the sensing channel
can be expressed as
\begin{equation}
\mathbf{G}=\rho_{s}\mathbf{f}_{r}\mathbf{f}_{t}^{H},
\end{equation}
where $\mathbf{f}_{t}=\left[e^{j\frac{2\pi}{\lambda}\|\mathbf{t}_{1}-\mathbf{s}\|},\dots,e^{j\frac{2\pi}{\lambda}\|\mathbf{t}_{N_{t}}-\mathbf{s}\|}\right]^{T}$
is the transmit near-field response vector, $\mathbf{f}_{r}=\left[e^{j\frac{2\pi}{\lambda}\|\mathbf{r}_{1}-\mathbf{s}\|},\dots,e^{j\frac{2\pi}{\lambda}\|\mathbf{r}_{N_{r}}-\mathbf{s}\|}\right]^{T}$
is the receive near-field response vector, and $\rho_{s}$ is the
round-trip channel coefficient. {Similar to \cite{liu2021cramer,hua2025near}, we assume that target tracking is employed whereby the radar aims to steer the sensing signal toward the previously estimated target position, and that the appropriate beamforming error is negligible.}

\subsection{System with Linear Precoding}
For the system with linear precoding, the transmitted ISAC
signal that ensures high-quality communication and sensing functionalities
is given by 
\begin{equation}
\mathbf{x}=\sum_{k=1}^{K}\mathbf{W}_{k}\mathbf{s}_{k}+\mathbf{v}s_{0},
\end{equation}
where $\mathbf{W}_{k}\in\mathbb{C}^{N_{t}\times N_{u}}$ is the communication
precoding matrix for transmission to user $k$, and $\mathbf{s}_{k}\in\mathbb{C}^{N_{u}\times1}$
is the corresponding transmitted signal, which consists of \ac{iid}
symbols that are distributed according to $\mathcal{CN}(0,1)$. Moreover,
$\mathbf{v}\in\mathbb{C}^{N_{t}\times1}$ is the sensing transmit
beamformer, and $s_{0}$ is the dedicated sensing signal that satisfies $\mathbb{E}\{\left|s_{0}\right|^{2}\}=1$ {and is statistically independent from the communication signals}.

\paragraph{Communication Model}

For user $k$, the received signal is given by
\begin{equation}
\mathbf{y}_{k}=\mathbf{H}_{k}\mathbf{W}_{k}\mathbf{s}_{k}+\mathbf{H}_{k}\sum_{u=1,u\neq k}^{K}\mathbf{W}_{u}\mathbf{s}_{u}+\mathbf{H}_{k}\mathbf{v}s_{0}+\mathbf{n}_{k},\label{eq:Rec_comm_sig}
\end{equation}
where $\mathbf{H}_{k}$ is given (1) and $\mathbf{n}_{k}\in\mathbb{C}^{N_{u}\times1}$ is the noise
vector with \ac{iid} elements distributed according to $\mathcal{CN}(0,\sigma_{k}^{2})$,
where $\sigma_{k}^{2}$ is the noise variance at user $k$.

From (\ref{eq:Rec_comm_sig}), the achievable rate for user $k$ is written as\footnote{{Note that the rates for both precoding schemes are expressed in nat/s here for mathematical convenience. On the other hand, these rates are expressed in bit/s in Section V. The rate expressed in bits/s is obtained by dividing the rate expressed in nat/s by ln(2).}}
\begin{align}
R_{L,k}= & \ln\Bigg|\mathbf{I}+\mathbf{H}_{k}\mathbf{W}_{k}\mathbf{W}_{k}^{H}\mathbf{H}_{k}^{H}\Biggl(\sum_{u=1,u\neq k}^{K}\mathbf{H}_{k}\mathbf{W}_{u}\mathbf{W}_{u}^{H}\mathbf{H}_{k}^{H}\nonumber \\
 & +\mathbf{H}_{k}\mathbf{v}\mathbf{v}^{H}\mathbf{H}_{k}^{H}+\sigma_{k}^{2}\mathbf{I}_{N_{u}}\Biggl)^{-1}\Bigg|.\label{eq:Rk_def-2}
\end{align}

Note that the communication rate in (\ref{eq:Rk_def-2}) is expressed
in nat/s for mathematical convenience.

\paragraph{Sensing Model}

At the receive ULA of the BS, the receive signal is given by 
\begin{equation}
\mathbf{y}_{B}=\mathbf{G}\mathbf{v}s_{0} +\mathbf{G}\sum_{u=1}^{K}\mathbf{W}_{u}\mathbf{s}_{u}+\mathbf{z},
\end{equation}
where $\mathbf{G}$ is given by (2) and $\mathbf{z}\in\mathbb{C}^{N_{r}\times1}$ is the noise vector
with i.i.d. elements distributed according to $\mathcal{CN}(0,\sigma_{z}^{2})$.

After implementing the receive signal combiner $\mathbf{u}\in\mathbb{C}^{N_{r}\times1}$,
the resulting signal is given by
\begin{equation}
y_{s}=\mathbf{u}^{H}\mathbf{y}_{B}=\mathbf{u}^{H}\mathbf{G}\mathbf{v}s_{0}+\mathbf{u}^{H}\mathbf{G}\sum_{k=1}^{K}\mathbf{W}_{k}\mathbf{s}_{k}+\mathbf{u}^{H}\mathbf{z},
\end{equation}
and the radar sensing capability can be evaluated via the sensing SINR,
which is defined as 
\begin{equation}
\gamma_{s}=\frac{P_{s}}{\text{\ensuremath{\mathbf{u}}}^{H}\left(\sum_{k=1}^{K}\mathbf{G}\mathbf{W}_{k}\mathbf{W}_{k}^{H}\mathbf{G}^{H}+\sigma_{z}^{2}\mathbf{I}_{N_{r}}\right)\text{\ensuremath{\mathbf{u}}}},\label{eq:SINR_def}
\end{equation}
where $P_{s}=\text{\ensuremath{\mathbf{u}}}^{H}\mathbf{G}\mathbf{v}\mathbf{v}^{H}\mathbf{G}^{H}\text{\ensuremath{\mathbf{u}}}$
is the sensing signal power.

\subsection{System with ZF Precoding}

In the case of ZF precoding, the transmitted ISAC signal is given
by 
\begin{equation}
\mathbf{x}=\mathbf{P}_{e}\mathbf{s}_{e}+\mathbf{v}s_{0},
\end{equation}
where $\mathbf{s}_{e}=[\mathbf{s}_{1}^{T},\mathbf{s}_{2}^{T},\dots,\mathbf{s}_{K}^{T}]^{T}\in\mathbb{C}^{KN_{u}\times1}$.
The precoding matrix that is defined as 
$
\mathbf{P}_{e}=\beta_{e}\mathbf{H}_{e}^{H}(\mathbf{H}_{e}\mathbf{H}_{e}^{H})^{-1}\in\mathbb{C}^{N_{t}\times KN_{u}}$,
where $\mathbf{H}_{e}=[\mathbf{H}^T_{1},\mathbf{H}^T_{2}.\dots,\mathbf{H}^T_{K}]^T\in\mathbb{C}^{KN_{u}\times N_{t}},$
 and the scaling coefficient 
\begin{equation}
\beta_{e}=\sqrt{\frac{P_{\max}}{\mathrm{Tr((\mathbf{H}_{e}\mathbf{H}_{e}^{H})^{-1})}}},
\end{equation}
{maintains a constant communication power budget $P_{\max}$}. {It should be noted that ZF precoding can be applied only when the condition $N_{t} \ge K N_{u}$
 is satisfied.
}

\paragraph{Communication Model}

From the previous expressions, a compact representation of the received
signal vectors for all users, $\mathbf{y}_{e}=[\mathbf{y}_{1}^{T},\mathbf{y}_{2}^{T}.\dots,\mathbf{y}_{K}^{T}]^{T}$,
can be expressed as
\begin{equation}
\mathbf{y}_{e}=\mathbf{H}_{e}\mathbf{x}+\mathbf{n}_{e}=\beta_{e}\mathbf{s}_{e}+\mathbf{H}_{e}\mathbf{v}s_{0}+\mathbf{n}_{e},
\end{equation}
where $\mathbf{n}_{e}=[\mathbf{n}_{1}^{T},\mathbf{n}_{2}^{T}.\dots,\mathbf{n}_{K}^{T}]^{T}$.
For user $k$, the received signal vector is given by 
\begin{equation}
\mathbf{y}_{k}=\beta_{e}\mathbf{s}_{k}+\mathbf{H}_{k}\mathbf{v}s_{0}+\mathbf{n}_{k},
\end{equation}
and the achievable rate for this user can be written as
\begin{align}
R_{Z,k}= & \ln\Bigg|\beta_{e}^{2}\mathbf{I}+\mathbf{H}_{k}\mathbf{v}\mathbf{v}^{H}\mathbf{H}_{k}^{H}+\sigma_{k}^{2}\mathbf{I}_{N_{u}}\Bigg|\nonumber \\
 & -\ln\Bigg|\mathbf{H}_{k}\mathbf{v}\mathbf{v}^{H}\mathbf{H}_{k}^{H}+\sigma_{k}^{2}\mathbf{I}_{N_{u}}\Bigg|.
\end{align}

\paragraph{Sensing Model}

The receive signal vector at the receive ULA of the BS is given by
\begin{equation}
\mathbf{y}_{B}=\mathbf{G}\mathbf{v}s_{0}+\mathbf{G}\mathbf{P}_{e}\mathbf{s}_{e}+\mathbf{z}.
\end{equation}

After implementing the receive signal combiner $\mathbf{u}\in\mathbb{C}^{N_{r}\times1}$,
the resulting signal is written as 
\begin{equation}
y_{s}=\mathbf{u}^{H}\mathbf{y}_{B}=\mathbf{u}^{H}\mathbf{G}\mathbf{v}s_{0} +\mathbf{u}^{H}\mathbf{G}\mathbf{P}_{e}\mathbf{s}_{e}+\mathbf{u}^{H}\mathbf{z},
\end{equation}
and the sensing SINR is given as 
\begin{equation}
\gamma_{s}=\frac{P_{s}}{\text{\ensuremath{\mathbf{u}}}^{H}\left(\mathbf{G}\mathbf{P}_{e}\mathbf{P}_{e}^{H}\mathbf{G}^{H}+\sigma_{z}^{2}\mathbf{I}_{N_{r}}\right)\text{\ensuremath{\mathbf{u}}}},\label{eq:SINR_def-1}
\end{equation}
where $P_{s}=\text{\ensuremath{\mathbf{u}}}^{H}\mathbf{G}\mathbf{v}\mathbf{v}^{H}\mathbf{G}^{H}\text{\ensuremath{\mathbf{u}}}$
is the sensing signal power.

\subsection{Problem Formulation}

In this paper, our goal is to maximize the WSR, while at the same
time maintaining a minimum \ac{SINR} performance level of $\gamma_{s}$ for
target sensing. Therefore, the appropriate optimization problem for
the system with linear precoding can be formulated as follows: 
\begin{subequations}
\label{eq:Orig_WSR_prob}
\begin{align}
&\text{\ensuremath{\underset{\mathbf{u},\{\mathbf{W}_{k}\},\mathbf{v},\{\mathbf{t}_{m}\},\{\mathbf{q}_{k,b}\}}{\mathrm{maximize}}}}  \;\text{WSR}=\sum_{k=1}^{K}w_{k}R_{L,k}\label{eq:WSR_obj}\\
&\hspace{2em}\text{s.t.} \quad \;\text{Tr}\left(\sum_{k=1}^{K}\mathbf{W}_{k}\mathbf{W}_{k}^{H}\right)\le P_{\max},\label{eq:Pmax}\\
 & \hspace{2.25em}\;\gamma_{s}\ge\gamma_{0},\label{eq:SINRmin}\\
 & \hspace{2.25em}\;{||\text{\ensuremath{\mathbf{v}}}||^{2} \le 1},\label{eq:TX_BF}\\
 & \hspace{2.25em}\;||\text{\ensuremath{\mathbf{u}}}||^{2}=1,\label{eq:RX_BF}\\
 &\hspace{2.25em} \;\mathbf{t}_m \in C_t, \;\forall m,\label{eq:BS_region}\\
 &\hspace{2.25em} \;\mathbf{q}_{k,b} \in C_k, \; \forall k, b,\label{eq:user_region}\\
 &\hspace{2.25em} \;||\mathbf{t}_{m_{1}}-\mathbf{t}_{m_{2}}||\ge d_{\min},1\le m_{1}\neq m_{2}\le N_{t},\label{eq:TX_space}\\
 &\hspace{2.25em} \;||\mathbf{q}_{k,b_{1}}-\mathbf{q}_{k,b_{2}}||\ge d_{\min},1\le b_{1}\neq b_{2}\le N_{u}, \text{\ensuremath{\forall}}k,\label{eq:User_space}
\end{align}
\end{subequations}
 where $w_{k}$ is the rate weight for user $k$.
It should be noted that constraint (\ref{eq:Pmax}) {specifies that the communication power budget} is $P_{\max}$. Constraint (\ref{eq:SINRmin}) specifies that
the sensing SINR has to be above the threshold $\gamma_{0}$.
Constraints (\ref{eq:TX_BF}) and (\ref{eq:RX_BF}) indicate the powers of
the sensing transmit beamformer and the sensing receive combiner. Constraints
(\ref{eq:BS_region}) and (\ref{eq:user_region})  specify the moving regions of
the transmit  and users' MAs, respectively. Constraints (\ref{eq:TX_space}) and (\ref{eq:User_space})
ensure the minimum inter-MA distance $d_{\min}$ for the transmit
and users' MAs.
Similarly as for linear precoding, the appropriate optimization problem for
the system with ZF precoding is expressed as 
\begin{subequations}
\label{eq:Orig_WSR_prob_ZF}
\begin{align}
&\text{\ensuremath{\underset{\mathbf{u},\mathbf{v},\{\mathbf{t}_{m}\},\{\mathbf{q}_{k,b}\}}{\mathrm{maximize}}}}  \;\mathrm{WSR}=\sum_{k=1}^{K}w_{k}R_{Z,k}\\
&\qquad\text{s.t.} \; {\mathrm{rank}(\mathbf{H}_{e})=KN_{u},} \label{eq:rank_cons}\\
 & \hspace{2em}(\ref{eq:SINRmin}), (\ref{eq:TX_BF}), (\ref{eq:RX_BF}), (\ref{eq:BS_region}), (\ref{eq:user_region}), (\ref{eq:TX_space}), (\ref{eq:User_space}) \nonumber\\
 & \hspace{5em} \mathrm{from} \, (\ref{eq:Orig_WSR_prob}) \nonumber.
\end{align}
\end{subequations}
{Constraint (\ref{eq:rank_cons}) ensures that the rank of $\mathbf{H}_{e}$ is always maintained, which is a necessary condition for implementation of ZF precoding.}  
An important difference between the previous two problems is that the
precoding matrices $\{\mathbf{W}_{k}\}$ are optimization variables
in (\ref{eq:Orig_WSR_prob}), while the precoding matrix $\mathbf{P}_{e}$
is not considered as an optimization variable in (\ref{eq:Orig_WSR_prob_ZF}).
The reason is that $\{\mathbf{W}_{k}\}$ are independent from other
optimization variables in (\ref{eq:Orig_WSR_prob}). On the other hand, $\mathbf{P}_{e}$
depends on the users' channel matrices and consequently on the positions of the BS and users' MAs.
Therefore, the optimization
of the positions of these MAs  will implicitly determine $\mathbf{P}_{e}$.

\section{Problem Solution for System with Linear Precoding}

This problem (\ref{eq:Orig_WSR_prob}) is difficult to solve due to
the non-convexity of the objective function, the coupling between
the optimization variables, and the unit modulus constraints. To deal
with it, we propose an AO-based algorithm which individually optimizes
each of the optimization variables separately and is elaborated in
more details in the sequel.

\subsection{Optimization of the Sensing Receive Beamformer}

Note that the optimal sensing receive beamformer $\mathbf{u}$ has
to maximize the sensing SINR $\gamma_{s}$ in (\ref{eq:SINR_def}).
After reformulating the numerator of $\gamma_{s}$ as
\begin{equation}
\text{\ensuremath{\mathbf{u}}}^{H}\mathbf{G}\mathbf{v}\mathbf{v}^{H}\mathbf{G}^{H}\text{\ensuremath{\mathbf{u}}}=(\rho_{s}^{2}\mathbf{f}_{t}^{H}\mathbf{v}\mathbf{v}^{H}\mathbf{f}_{t})\times(\text{\ensuremath{\mathbf{u}}}^{H}\mathbf{f}_{r}\mathbf{f}_{r}^{H}\text{\ensuremath{\mathbf{u}}}),
\end{equation}
we can observe that $\rho_{s}^{2}\mathbf{f}_{t}^{H}\mathbf{v}\mathbf{v}^{H}\mathbf{f}_{t}\ge0$
is independent of $\text{\ensuremath{\mathbf{u}}}$. Hence, the appropriate
optimization problem is expressed as 
\begin{subequations}
\label{eq:Opt_prob-u}
\begin{align}
\underset{\mathbf{u}}{\mathrm{maximize}} & \;\frac{\text{\ensuremath{\mathbf{u}}}^{H}\mathbf{f}_{r}\mathbf{f}_{r}^{H}\text{\ensuremath{\mathbf{u}}}}{\text{\ensuremath{\mathbf{u}}}^{H}\mathbf{D}_{l}\text{\ensuremath{\mathbf{u}}}}\\
\text{s.t.} & \;(\ref{eq:RX_BF}),\nonumber 
\end{align}
\end{subequations}
 where $\mathbf{D}_{l}=\sum_{u=1}^{K}\mathbf{G}\mathbf{W}_{u}\mathbf{W}_{u}^{H}\mathbf{G}^{H}+\sigma_{z}^{2}\mathbf{I}_{N_r}$.
From \cite{he2023full}, the optimal $\mathbf{u}$ is given by 

\begin{equation}
\text{\ensuremath{\mathbf{u}}}^{*}=\frac{\mathbf{D}_{l}^{-1}\mathbf{f}_{r}}{||\mathbf{D}_{l}^{-1}\mathbf{f}_{r}||}.\label{eq:u_opt}
\end{equation}

\subsection{Optimization of the Communication Precoding Matrices}

We remark that the objective function in (\ref{eq:WSR_obj}) is neither convex nor
concave with respect to the precoding matrices $\{\mathbf{W}_{k}\}$.
To deal with this, we exploit the following inequality to derive a
tight concave lower bound on the achievable rate of each user. For matrices $\mathbf{X}$
and $\mathbf{\bar{X}}$ with size $p\times q$, and matrices $\mathbf{Y}\succcurlyeq\mathbf{0}$
and $\mathbf{\bar{Y}}\succcurlyeq\mathbf{0}$ with size $p\times p$,
the following inequality is valid \cite{zhang2023discerning}:
\begin{align}
&\ln\left|\mathbf{I}+\mathbf{X}^{H}\mathbf{Y}^{-1}\mathbf{X}\right|\ge\ln\left|\mathbf{I}+\mathbf{\bar{X}}^{H}\bar{\mathbf{Y}}^{-1}\bar{\mathbf{X}}\right|\nonumber \\
&\quad -\text{Tr}\{\mathbf{\bar{X}}^{H}\bar{\mathbf{Y}}^{-1}\bar{\mathbf{X}}\}+2\mathfrak{R}\{\text{Tr}\{\mathbf{\bar{X}}^{H}\bar{\mathbf{Y}}^{-1}\mathbf{X}\}\}\nonumber \\
&\qquad -\text{Tr}\{(\bar{\mathbf{Y}}+\mathbf{\bar{X}}\mathbf{\bar{X}}^{H})^{-1}\mathbf{\bar{X}}\mathbf{\bar{X}}^{H}\bar{\mathbf{Y}}^{-1}(\mathbf{Y}+\mathbf{X}\mathbf{X}^{H})\}. \label{eq:Rk_LB}
\end{align}

Let $\{\mathbf{W}_{k}^{(n)}\}$ denote the value of $\{\mathbf{W}_{k}\}$
after $n$ iterations, $\mathbf{X}=\mathbf{H}_{k}\mathbf{W}_{k}$,
$\bar{\mathbf{X}}=\mathbf{H}_{k}\mathbf{W}_{k}^{(n)}$, $\mathbf{Y}=\sum_{u=1,u\neq k}^{K}\mathbf{H}_{k}\mathbf{W}_{u}\mathbf{W}_{u}^{H}\mathbf{H}_{k}^{H}+\mathbf{H}_{k}\mathbf{v}\mathbf{v}^{H}\mathbf{H}_{k}^{H}+\sigma_{k}^{2}\mathbf{I}_{N_{u}}$
and $\bar{\mathbf{Y}}=\sum_{u=1,u\neq k}^{K}\mathbf{H}_{k}\mathbf{W}_{u}^{(n)}(\mathbf{W}_{u}^{(n)})^{H}\mathbf{H}_{k}^{H}+\mathbf{H}_{k}\mathbf{v}\mathbf{v}^{H}\mathbf{H}_{k}^{H}+\sigma_{k}^{2}\mathbf{I}_{N_{u}}$.
After a few simple mathematical steps, (\ref{eq:Rk_LB}) results in
\begin{align}
R_{L,k}  \ge &\hat{R}_{L,k}=\ln\left|\mathbf{I}_{N_u}+\mathbf{W}_{k}^{(n)H}\mathbf{H}_{k}^{H}(\mathbf{F}_{k}^{(n)})^{-1}\mathbf{H}_{k}\mathbf{W}_{k}^{(n)}\right|\nonumber \\
 & -\text{Tr}(\mathbf{W}_{k}^{(n)H}\mathbf{H}_{k}^{H}(\mathbf{F}_{k}^{(n)})^{-1}\mathbf{H}_{k}\mathbf{W}_{k}^{(n)})\nonumber \\
 & +2\mathfrak{R}(\text{Tr}(\mathbf{W}_{k}^{(n)H}\mathbf{H}_{k}^{H}(\mathbf{F}_{k}^{(n)})^{-1}\mathbf{H}_{k}\mathbf{W}_{k}))\nonumber \\
 & -\text{Tr}(\sum_{u=1}^{K}\mathbf{W}_{u}^{H}\mathbf{H}_{k}^{H}\mathbf{A}_{k}\mathbf{H}_{k}\mathbf{W}_{u})\nonumber \\
 & -\text{Tr}(\mathbf{A}_{k}(\mathbf{H}_{k}\mathbf{v}\mathbf{v}^{H}\mathbf{H}_{k}^{H}+\sigma_{k}^{2}\mathbf{I}_{N_{u}})),
\end{align}
where $\mathbf{F}_{k}^{(n)}=\bar{\mathbf{Y}}$ and $\mathbf{A}_{k}=(\mathbf{F}_{k}^{(n)}+\mathbf{H}_{k}\mathbf{W}_{k}^{(n)}\mathbf{W}_{k}^{(n)H}\mathbf{H}_{k}^{H})^{-1}\mathbf{H}_{k}\mathbf{W}_{k}^{(n)}\mathbf{W}_{k}^{(n)H}\mathbf{H}_{k}^{H}(\mathbf{F}_{k}^{(n)})^{-1}$.

Now, it is easy to see that this lower bound is a concave function
of $\{\mathbf{W}_{k}\}$. Furthermore, the constraint (\ref{eq:SINRmin})
can be reformulated as 
\begin{gather}
\kappa_{l}\!\triangleq\!\mathbf{u}^{H}\mathbf{G}(\gamma_{0}\sum_{u=1}^{K}\!\mathbf{W}_{u}\mathbf{W}_{u}^{H}-\mathbf{v}\mathbf{v}^{H})\mathbf{G}^{H}\mathbf{u}+\gamma_{0}\sigma_{z}^{2}\mathbf{u}^{H}\mathbf{u}\le0\!.\label{eq:SINR_ref}
\end{gather}

Hence, the precoding matrix optimization problem can be formulated
as 
\begin{subequations}
\label{eq:Prec_opt_prob}
\begin{align}
\underset{\{\mathbf{W}_{k}\}}{\mathrm{maximize}} & \;\widehat{\mathrm{WSR}}=\sum_{k=1}^{K}w_{k}\hat{R}_{L,k}\label{eq:WSR_hat}\\
\text{s.t.} & \;\text{Tr}(\sum_{k=1}^{K}\mathbf{W}_{k}\mathbf{W}_{k}^{H})\le P_{\max},\label{eq:Pmax-1-2}\\
 & \;(\ref{eq:SINR_ref}),\nonumber 
\end{align}
\end{subequations}
which can be solved by any conventional optimization solver. Finally,
the algorithm for optimizing the precoding matrices $\{\mathbf{W}_{k}\}$
is summarized in Algorithm \ref{alg:Prec_mat_opt}. 
\begin{algorithm}[t]
\small
\caption{SCA-based Method for Optimizing the Communication Precoding Matrices \label{alg:Prec_mat_opt}}
\SetAlgoNoLine
\DontPrintSemicolon
\LinesNumbered 

\KwIn{ $\{\mathbf{W}_{k}^{(0)}\}$, $n\leftarrow0$}

\Repeat{$\widehat{\mathrm{WSR}}^{(n)}-\widehat{\mathrm{WSR}}^{(n-1)}<\epsilon_{s}$}{

Solve (\ref{eq:Prec_opt_prob}) to obtain $\{\mathbf{W}_{k}^{(n+1)}\}$
\;

Calculate $\widehat{\mathrm{WSR}}^{(n+1)}$ according to (\ref{eq:WSR_hat})\;

$n\leftarrow n+1$ \;

}

\KwOut{ $\{\mathbf{W}_{k}^{*}\}=\{\mathbf{W}_{k}^{(n)}\}$ }
\end{algorithm}

\subsection{Optimization of the Sensing Transmit Beamformer}

Similarly as in the previous subsection, we use the SCA method
to find the optimal transmit beamformer $\mathbf{v}$, but we now introduce the covariance
matrix $\mathbf{V}=\mathbf{v}\mathbf{v}^{H}$. As a result, the achievable
rate of user $k$ can be written as
\begin{align}
&R_{L,k}  \!=\!\ln\Bigg|\sum_{u=1}^{K}\mathbf{H}_{k}\mathbf{W}_{u}\mathbf{W}_{u}^{H}\mathbf{H}_{k}^{H}+\mathbf{H}_{k}\mathbf{V}\mathbf{H}_{k}^{H}+\sigma_{k}^{2}\mathbf{I}_{N_{u}}\Bigg|\nonumber \\
 & -\ln\Bigg|\sum_{u=1,u\neq k}^{K}\!\!\mathbf{H}_{k}\mathbf{W}_{u}\mathbf{W}_{u}^{H}\mathbf{H}_{k}^{H}\!+\!\mathbf{H}_{k}\mathbf{V}\mathbf{H}_{k}^{H}\!+\!\sigma_{k}^{2}\mathbf{I}_{N_{u}}\Bigg|\label{eq:Rk_for_v}.
\end{align}

Denoting by $\mathbf{B}_{k}(\mathbf{V})=\sum_{u=1,u\neq k}^{K}\mathbf{H}_{k}\mathbf{W}_{u}\mathbf{W}_{u}^{H}\mathbf{H}_{k}^{H}+\mathbf{H}_{k}\mathbf{V}\mathbf{H}_{k}^{H}+\sigma_{k}^{2}\mathbf{I}_{N_{u}}$,
the first-order Taylor approximation to the second term in (\ref{eq:Rk_for_v}) is given by
\begin{align}
\ln\left|\mathbf{B}_{k}(\mathbf{V})\right| & \le\ln\left|\mathbf{B}_{k}(\mathbf{V}^{(n)})\right|\nonumber \\
 & +\text{Tr}(\mathbf{H}_{k}^{H}(\mathbf{B}_{k}(\mathbf{V}^{(n)}))^{-1}\mathbf{H}_{k}(\mathbf{V}-\mathbf{V}^{(n)})),
\end{align}
where $\mathbf{V}^{(n)}$ is the value of $\mathbf{V}$ after $n$
iterations. Now, (\ref{eq:Rk_for_v}) can be lower bounded as 
\begin{align}
&R_{L,k}  \ge\bar{R}_{L,k}=\ln\left|\mathbf{B}_{k}(\mathbf{V})+\mathbf{H}_{k}\mathbf{W}_{k}\mathbf{W}_{k}^{H}\mathbf{H}_{k}^{H}\right|\nonumber \\
 &\;\; -\ln\left|\mathbf{B}_{k}(\mathbf{V}^{(n)})\right|-\text{Tr}(\mathbf{H}_{k}^{H}(\mathbf{B}_{k}(\mathbf{V}^{(n)}))^{-1}\mathbf{H}_{k}(\mathbf{V}-\mathbf{V}^{(n)})).
\end{align}

From the above, the appropriate optimization problem can be formulated as 
\begin{subequations}
\label{eq:Opt_prec_vest}
\begin{align}
\underset{\mathbf{V}}{\mathrm{maximize}} & \;\overline{\mathrm{WSR}}=\sum_{k=1}^{K}w_{k}\bar{R}_{L,k}\label{eq:WSR_bar}\\
\text{s.t.} & \;\text{Tr}(\mathbf{V})\le1,\label{eq:Tr_V}\\
 & \;\text{rank}(\mathbf{V})=1,\label{eq:Rank1}\\
 & \;(\ref{eq:SINR_ref}).\nonumber 
\end{align}
\end{subequations}
 We observe that the rank-1 constraint (\ref{eq:Rank1}) is not convex.
Also, this constraint is equivalent to the maximization of $\beta_{\max}(\mathbf{V})-\text{Tr}(\mathbf{V})$,
where $\beta_{\max}(\mathbf{V})$ is the largest eigenvalue of $\mathbf{V}$.
Exploiting the inequality $\beta_{\max}(\mathbf{V})\ge\beta_{\max}(\mathbf{V}^{(n)})+\text{Tr}(\boldsymbol{\chi}^{(n)}\boldsymbol{\chi}^{(n)H}(\mathbf{V}-\mathbf{V}^{(n)}))$,
where $\boldsymbol{\chi}^{(n)}$ is the eigenvector corresponding
to $\beta_{\max}(\mathbf{V}^{(n)})$, the previous optimization problem
can be expressed as 
\begin{subequations}
\label{eq:Opt_sense_prec_vec}
\begin{align}
\underset{\mathbf{V}}{\mathrm{maximize}} & \;\overline{\mathrm{WSR}}+\zeta\mathcal{M}\\
\text{s.t.} & \;(\ref{eq:SINR_ref}),(\ref{eq:Tr_V}),\nonumber 
\end{align}
\end{subequations}
where $\mathcal{M}=\beta_{\max}(\mathbf{V}^{(n)})+\text{Tr}(\boldsymbol{\chi}^{(n)}\boldsymbol{\chi}^{(n)H}(\mathbf{V}-\mathbf{V}^{(n)}))-\text{Tr}(\mathbf{V})$,
and $\zeta$ is the penalty parameter. We notice that this optimization
problem is now convex, and can thus be solved by a standard optimization
solver. The algorithm for optimizing the transmit sensing beamformer
$\mathbf{v}$ is presented in Algorithm \ref{alg:Prec_sense_vector}.
\begin{algorithm}[t]
\small
\caption{SCA-based Method for Optimizing the Sensing Transmit Beamformer \label{alg:Prec_sense_vector}}
\SetAlgoNoLine
\DontPrintSemicolon
\LinesNumbered 

\KwIn{ $\mathbf{v}$, $\zeta$, $n\leftarrow0$}

$\mathbf{V}^{(0)}=\mathbf{v}\mathbf{v}^{H}$

\Repeat{$\overline{\mathrm{WSR}}^{(n)}-\overline{\mathrm{WSR}}^{(n-1)}<\epsilon_{s}$}{

Calculate $\beta_{\max}(\mathbf{V}^{(n)})$ and $\boldsymbol{\chi}^{(n)}$

Solve (\ref{eq:Opt_sense_prec_vec}) to obtain $\mathbf{V}^{(n+1)}$
\;

Calculate $\overline{\mathrm{WSR}}^{(n+1)}$ according to (\ref{eq:WSR_bar})\;

$n\leftarrow n+1$ \;

}

Calculate $\beta_{\max}(\mathbf{V}^{(n)})$ and $\boldsymbol{\chi}^{(n)}$

\KwOut{ $\mathbf{v}^{*}=(\beta_{\max}(\mathbf{V}^{(n)}))^{1/2}\boldsymbol{\chi}^{(n)}$
}
\end{algorithm}

\subsection{Optimization of MA Positions of User $k$}

The optimization of the MA positions of user $k$ aims to maximize
its achievable communication rate, $R_{L,k}$. The appropriate
optimization problem can be formulated as  
\begin{subequations}
\label{eq:Opt_qk}
\begin{align}
\underset{\{\mathbf{q}_{k,b}\}}{\mathrm{maximize}} & \;R_{L,k}\\
\text{s.t.} & \;(\ref{eq:user_region}),(\ref{eq:User_space}).\label{eq:Tr_V-1}
\end{align}
\end{subequations}
 To solve this problem, we utilize the PGM approach. If the MA positions of user $k$ after 
 $n$ iterations are denoted by $\mathbf{q}_{k}^{(n)}$, then the MA
positions in the $(n+1)$-th iteration can be obtained as
\begin{equation}
\mathbf{q}_{k}^{(n+1)}=P(\mathbf{q}_{k}^{(n)}+\mu_{k}^{(n)}\nabla_{\mathbf{q}_{k}}R_{k}(\mathbf{q}_{k}^{(n)})),
\end{equation}
where $\mu_{k}^{(n)}$ is the step size. The gradient of $R_{k}$
\ac{wrt} $\mathbf{q}_{k}$ is determined by the gradients of $R_{k}$
\ac{wrt} the individual MA coordinates:
\begin{align*}
\nabla_{\mathbf{q}_{k}}R_{L,k} & =[\nabla_{x_{k,1}}R_{L,k},\nabla_{y_{k,1}}R_{k},0,\dots,\\
 & \nabla_{x_{k,N_{u}}}R_{L,k},\nabla_{y_{k,N_{u}}}R_{L,k},0]^{T}.
\end{align*}
and these gradients are provided in the following lemma.
\begin{lem}
\label{lem:Proof_lem_MA_user}The gradients of $R_{k}$ \ac{wrt}
the x and y coordinate of the b-th MA of user $k$ are given by (\ref{eq:dRk_x}) 
and (\ref{eq:dRk_y}) on the next page, where 
\begin{align}
\mathbf{C}_{1} & =\mathbf{T}_{1}\mathbf{H}_{k}^{H}\mathbf{A}_{1}^{-1},\\
\mathbf{C}_{2,k} & =\mathbf{T}_{2}\mathbf{H}_{k}^{H}\mathbf{A}_{2,k}^{-1},
\end{align}
where $\mathbf{T}_{1}=\sum_{u=1}^{K}\mathbf{W}_{u}\mathbf{W}_{u}^{H}+\mathbf{v}\mathbf{v}^{H},\mathbf{T}_{2,k}=\mathbf{T}_{1}-\mathbf{W}_{k}\mathbf{W}_{k}^{H}$,
$\mathbf{A}_{1}=\mathbf{H}_{k}\mathbf{T}_{1}\mathbf{H}_{k}^{H}+\sigma_{k}^{2}\mathbf{I}_{N_{u}}$
and $\mathbf{A}_{2,k}=\mathbf{H}_{k}\mathbf{T}_{2,k}\mathbf{H}_{k}^{H}+\sigma_{k}^{2}\mathbf{I}_{N_{u}}$.
\end{lem}
\begin{IEEEproof}
See Appendix A. 
\begin{figure*}[tbh]
\begin{align}
\nabla_{x_{k,b}}R_{L,k} & =\frac{4\pi\rho_{k}}{\lambda}\mathfrak{I}\Big(\sum_{m=1}^{N_{t}}[\mathbf{C}_{2,k}(m,b)-\mathbf{C}_{1}(m,b)]\frac{x_{k,b}-x_{m}}{||\mathbf{t}_{m}-\mathbf{q}_{k,b}||}e^{j\frac{2\pi}{\lambda}||\mathbf{t}_{m}-\mathbf{q}_{k,b}||}\Big)\label{eq:dRk_x}\\
\nabla_{y_{k,b}}R_{L,k} & =\frac{4\pi\rho_{k}}{\lambda}\mathfrak{I}\Big(\sum_{m=1}^{N_{t}}[\mathbf{C}_{2,k}(m,b)-\mathbf{C}_{1}(m,b)]\frac{y_{k,b}-y_{m}}{||\mathbf{t}_{m}-\mathbf{q}_{k,b}||}e^{j\frac{2\pi}{\lambda}||\mathbf{t}_{m}-\mathbf{q}_{k,b}||}\Big)\label{eq:dRk_y}
\end{align}
\hrule
\vspace{-1.5em}
\end{figure*}
\end{IEEEproof}
 The gradient projection is performed element-wise and independently
for each MA coordinate. For the $x$-coordinate of MA $b$, the gradient
projection is given by
\begin{equation}
P(x_{k,b})=\begin{cases}
x_{\max} & x_{\max}<x_{k,b},\\
x_{k,b} & x_{\min}\le x_{k,b}\le x_{\max},\\
x_{\min} & x_{k,b}<x_{\min},
\end{cases}\label{eq:Grad_proj}
\end{equation}
where $x_{\max}=x_{u,k}+a_{k}/2$ and $x_{\min}=x_{u,k}-a_{k}/2$.
The projection for $y$ coordinates is done in a similar way. In each
iteration, the step size $\mu_{k}^{(n)}$ is adjusted, i.e., decreased,
using the backtracking line search until the constraint (\ref{eq:User_space})
and the Armijo–Goldstein condition
\begin{equation}
R_{k}(\mathbf{q}_{k}^{(n+1)})-R_{k}(\mathbf{q}_{k}^{(n)})\ge\delta||\mathbf{q}_{k}^{(n+1)}-\mathbf{q}_{k}^{(n)}||^{2},
\end{equation}
where $\delta$ is a small positive number, are both satisfied.

\subsection{Optimization of the Transmit MA Positions at the BS \label{subsec:ZF-Prec-Opt-BS-TX}}

The optimal transmit MA positions at the BS are those that solve the following
optimization problem: 
\begin{subequations}
\begin{align}
\text{\ensuremath{\underset{\{\mathbf{t}_{m}\}}{\mathrm{maximize}}}} & \;\mathrm{WSR}=\sum_{k=1}^{K}w_{k}R_{L,k}\\
\text{s.t.} & \;(\ref{eq:SINRmin}),(\ref{eq:BS_region}),(\ref{eq:TX_space}).\label{eq:SINRmin-1}
\end{align}
\end{subequations}
 Using the equivalent formulation of (\ref{eq:SINRmin}) in (\ref{eq:SINR_ref})
and implementing the \ac{ALM} \cite{le2022algames}, we can reformulate
the previous optimization problem as 
\begin{subequations}
\label{eq:Lagr}
\begin{align}
\text{\ensuremath{\underset{\{\mathbf{t}_{m}\}}{\mathrm{minimize}}}} & \;L(\mathbf{t})=-\sum_{k=1}^{K}w_{k}R_{L,k}+\eta\kappa_{l}+\frac{1}{2}p\kappa_{l}^{2}\label{eq:Lt_obj}\\
\text{s.t.} & \;(\ref{eq:BS_region}),(\ref{eq:TX_space}),\label{eq:SINRmin-1-2}
\end{align}
\end{subequations}
 where $\mathbf{t} = [\mathbf{t}^T_1,\mathbf{t}^T_2,\dots,\mathbf{t}^T_{N_t}]^T$ and 
 $\eta$ is the Lagrangian multiplier and the penalty factor
is given by 
\begin{equation}
p=\begin{cases}
0 & \kappa_{l}<0
\:\text{and}\:\eta=0,
\\
p_{0} & \text{otherwise}.
\end{cases}
\end{equation}
To solve (\ref{eq:Lagr}), we implement the PGM via 
\begin{equation}
\mathbf{t}^{(n+1)}=P(\mathbf{t}^{(n)}-\nu^{(n)}\nabla_{\mathbf{t}}L(\mathbf{t}^{(n)})),
\end{equation}
where $\nu^{(n)}$ is the step size, and the gradient projection is
performed similarly as in (\ref{eq:Grad_proj}). The gradient $\nabla_{\mathbf{t}}L(\mathbf{t}^{(n)})$
is determined by the gradients of $L(\mathbf{t}^{(n)})$ w.r.t. the
coordinates of individual BS transmit MAs. For the $m$-th
BS transmit MA, these gradients are given by 
\begin{align}
\nabla_{x_{m}}L(\mathbf{t}) & =-\sum_{k=1}^{K}w_{k}\nabla_{x_{m}}R_{L,k}+(\eta+p\kappa_{l})\nabla_{x_{m}}\kappa_{l},\\
\nabla_{y_{m}}L(\mathbf{t}) & =-\sum_{k=1}^{K}w_{k}\nabla_{y_{m}}R_{L,k}+(\eta+p\kappa_{l})\nabla_{y_{m}}\kappa_{l},
\end{align}
where the gradients of $R_{L,k}$ and $\kappa_{l}$ w.r.t. the specified
MA coordinates are defined in the following lemma.

\begin{algorithm}[t]
\small
\caption{ALM for Optimizing the BS Transmit MA Positions\label{alg:ALM}}

\DontPrintSemicolon
\LinesNumbered 

\KwIn{$\mathbf{t}^{(0)}$, $\eta\leftarrow0$, $p_{0}$, $\theta>1$}

$i\leftarrow0$\;

\Repeat{$|\mathrm{WSR}(\mathbf{t}_{\text{\ensuremath{\mathrm{c}} }}^{(i)})-\mathrm{WSR}(\mathbf{t}_{\text{\ensuremath{\mathrm{c}} }}^{(i-1)})|<\epsilon_{f}$}{

Compute $L(\mathbf{t}^{(0)})$ according to (\ref{eq:Lt_obj})

$n\leftarrow0$\;
\Repeat{$\left|\frac{L(\mathbf{t}^{(n)})-L(\mathbf{t}^{(n-1)})}{L(\mathbf{t}^{(n)})}\right|<\epsilon_{L}$}{

\Repeat{$L(\mathbf{t}^{(n)})-L(\mathbf{t}^{(n+1)})\ge\delta||\mathbf{t}^{(n+1)}-\mathbf{t}^{(n)}||^{2}$
$\mathrm{\mathbf{and}}$ $||\mathbf{t}_{m_{1}}^{(n+1)}-\mathbf{t}_{m_{2}}^{(n+1)}||\ge d_{\min}$
$\:(\mathrm{for}\:\mathrm{all}\:m_{1}\neq m_{2})$}{

$\mathbf{t}^{(n+1)}=P(\mathbf{t}^{(n)}-\nu^{(n)}\nabla_{\mathbf{t}}L(\mathbf{t}^{(n)}))$\;

\If{$L(\mathbf{t}^{(n)})-L(\mathbf{t}^{(n+1)})<\delta||\mathbf{t}^{(n+1)}-\mathbf{t}^{(n)}||^{2}$
$\boldsymbol{\mathbf{\mathrm{or}}}$ $||\mathbf{t}_{m_{1}}^{(n+1)}-\mathbf{t}_{m_{2}}^{(n+1)}||<d_{\min}$
($\mathrm{for}\:\mathrm{any}\:m_{1}\neq m_{2}$)}{

$\nu^{(n)}\leftarrow\tau\nu^{(n)}$\;

}}

$n\leftarrow n+1$\;

}

$\mathbf{t}_{\text{c}}^{(i+1)}=\mathbf{t}^{(n)}$\;

$\mathbf{t}^{(0)}=\mathbf{t}^{(n)}$\;

$\eta=\max(0,\eta+p_{0}\kappa_{l})$\;

$p_{0}=\theta p_{0}$\; \label{penalty_p0}

$i\leftarrow i+1$

}

\KwOut{$\mathbf{t}^{*}=\mathbf{t}_{\text{c}}^{(i)}$}
\end{algorithm}

\begin{lem}
The gradients of $R_{L,k}$ w.r.t. the coordinates of the m-th BS
transmit MA are provided in (\ref{eq:dRk_xm}) and (\ref{eq:dRk_ym}), shown on the next page.
\begin{figure*}[tbh]
\begin{align}
\nabla_{x_{m}}R_{L,k} & =\frac{4\pi\rho_{k}}{\lambda}\mathfrak{I}\Big(\sum_{b=1}^{N_{u}}[\mathbf{C}_{2,k}(m,b)-\mathbf{C}_{1,k}(m,b)]\frac{x_{m}-x_{k}^{b}}{\left\Vert \mathbf{t}_{m}-\mathbf{q}_{k}^{b}\right\Vert }e^{j\frac{2\pi}{\lambda}\left\Vert \mathbf{t}_{m}-\mathbf{q}_{k}^{b}\right\Vert }\Big)\label{eq:dRk_xm}\\
\nabla_{y_{m}}R_{L,k} & =\frac{4\pi\rho_{k}}{\lambda}\mathfrak{I}\Big(\sum_{b=1}^{N_{u}}[\mathbf{C}_{2,k}(m,b)-\mathbf{C}_{1,k}(m,b)]\frac{y_{m}-y_{k}^{b}}{\left\Vert \mathbf{t}_{m}-\mathbf{q}_{k}^{b}\right\Vert }e^{j\frac{2\pi}{\lambda}\left\Vert \mathbf{t}_{m}-\mathbf{q}_{k}^{b}\right\Vert }\Big)\label{eq:dRk_ym}
\end{align}
\hrule
\end{figure*}
\end{lem}

\begin{IEEEproof}
The derivation of the gradients of $R_{L,k}$ \ac{wrt} to the BS
MA coordinates is similar to the derivation of the gradients of $R_{k}$
\ac{wrt} to the users' MA coordinates in Appendix A\ref{sec:Grad_qk_Rk}.
After substituting
\begin{align}
\frac{\text{d}\mathbf{H}_{k}(b,m)}{dx_{m}} & =j\frac{2\pi\rho_{k}}{\lambda}\frac{x_{m}-x_{k,b}}{||\mathbf{t}_{m}-\mathbf{q}_{k,b}||}e^{j\frac{2\pi}{\lambda}||\mathbf{t}_{m}-\mathbf{q}_{k,b}||},\label{eq:dHk_x-1}\\
\frac{\text{d}\mathbf{H}_{k}(b,m)}{dy_{m}} & =j\frac{2\pi\rho_{k}}{\lambda}\frac{y_{m}-y_{k,b}}{||\mathbf{t}_{m}-\mathbf{q}_{k,b}||}e^{j\frac{2\pi}{\lambda}||\mathbf{t}_{m}-\mathbf{q}_{k,b}||},\label{eq:dHk_y-1}
\end{align}
into (\ref{eq:dA1}) and (\ref{eq:dA2}), and a few simple algebraic
steps, we obtain (\ref{eq:dRk_xm}) and (\ref{eq:dRk_ym}).
\end{IEEEproof}

\begin{lem}
\label{lem:Proof_lem_MA_TX}The gradients of $\kappa_{l}$ w.r.t.
the coordinates of the m-th BS transmit MA are given by 
\begin{align}
\nabla_{x_{m}}\kappa_{l} & =-\frac{4\pi\rho_{s}}{\lambda}\mathfrak{I}\Big(\mathbf{e}(m)e^{j\frac{2\pi}{\lambda}\left\Vert \mathbf{t}_{m}-\mathbf{s}_{l}\right\Vert }\frac{x_{m}-x_{l}}{\text{\ensuremath{\left\Vert \mathbf{t}_{m}-\mathbf{s}_{l}\right\Vert }}}\Big),\label{eq:dh_xm}\\
\nabla_{y_{m}}\kappa_{l} & =-\frac{4\pi\rho_{s}}{\lambda}\mathfrak{I}\Big(\mathbf{e}(m)e^{j\frac{2\pi}{\lambda}\left\Vert \mathbf{t}_{m}-\mathbf{s}_{l}\right\Vert }\frac{y_{m}-y_{l}}{\text{\ensuremath{\left\Vert \mathbf{t}_{m}-\mathbf{s}_{l}\right\Vert }}}\Big),\label{eq:dh_ym}
\end{align}
for $\mathbf{M}=\gamma_{0}\sum_{u=1}^{K}\mathbf{W}_{u}\mathbf{W}_{u}^{H}-\mathbf{v}\mathbf{v}^{H}$
and $\mathbf{e}=\mathbf{f}_{r}^{H}\mathbf{u}\mathbf{u}^{H}\mathbf{G}\mathbf{M}$.
\end{lem}
\begin{IEEEproof}
See Appendix B\ref{sec:Grad_kz_mth_MA}.
\end{IEEEproof}
The overall algorithm for optimizing the BS transmit MA positions
is outlined in Algorithm \ref{alg:ALM}. In the first part, the MA
positions are iteratively optimized using the PGM until the convergence\footnote{It should be noted that this convergence criterion for $L(\mathbf{t})$ is based on the
relative change and not on the absolute change between two of its consecutive
values. The reason is that $L(\mathbf{t})$ can change by more than an order of magnitude during 
the optimization process, primarily
because of the change of the parameter $p_0$ (line \ref{penalty_p0}
in Algorithm \ref{alg:ALM}).} of $L(\mathbf{t})$, and these MA positions are saved as $\mathbf{t}_{\text{\ensuremath{\mathrm{c}}}}$.
Afterwards, the Lagrangian multiplier $\eta$ is recomputed and the parameter $p_{0}$,
that determines the value of the penalty factor $p$, is scaled by the constant $\theta$.
These steps are repeated until the convergence of the WSR, i.e., until
the difference between the WSRs calculated for the MA positions determined
by $\mathbf{t}_{\text{\ensuremath{\mathrm{c}} }}^{(i-1)}$ and $\mathbf{t}_{\text{\ensuremath{\mathrm{c}} }}^{(i)}$
is less than the specified threshold.

\subsection{Overall Algorithm}

\begin{algorithm}[t]
\small
\caption{Overall Optimization Algorithm for Solving (\ref{eq:Orig_WSR_prob})
\label{alg:Overall_alg}}
\DontPrintSemicolon
\LinesNumbered 

\KwIn{$\mathbf{u}^{(0)}$, $\{\mathbf{W}_{k}^{(0)}\}$, $\mathbf{v}^{(0)}$,
$\{\mathbf{q}_{k}^{(0)}\}$, $\mathbf{t}^{(0)}$, $\{\rho_{k}\}$,
$\rho_{s}$, $\{\mu_{k}^{(0)}\}$, $d_{\min}$, $\eta\leftarrow0$,
$\tau\in(0,1)$, $\delta>0$, $n\leftarrow0$}

\Repeat{convergence of $\mathrm{WSR}$ in (\ref{eq:WSR_obj})}{

Obtain $\mathbf{u}^{(n+1)}$ from (\ref{eq:u_opt})

Obtain $\{\mathbf{W}_{k}^{(n+1)}\}$ by Algorithm \ref{alg:Prec_mat_opt}\;

Obtain $\mathbf{v}^{(n+1)}$ by Algorithm \ref{alg:Prec_sense_vector}\;

\For{$k=1$ $\mathrm{to}$ $K$}{

\Repeat{$R_{L,k}(\mathbf{q}_{k}^{(n+1)})-R_{L,k}(\mathbf{q}_{k}^{(n)})\ge\delta||\mathbf{q}_{k}^{(n+1)}-\mathbf{q}_{k}^{(n)}||^{2}$
$\mathrm{\mathbf{and}}$ $||\mathbf{q}_{k,b_{1}}^{(n+1)}-\mathbf{q}_{k,b_{2}}^{(n+1)}||\ge d_{\min}\:(\mathrm{for}\:\mathrm{all}\:b_{1}\neq b_{2})$}{

$\mathbf{q}_{k}^{(n+1)}=P(\mathbf{q}_{k}^{(n)}+\mu^{(n)}\nabla_{\mathbf{q}_{k}}R_{k})$\;

\If{$R_{k}(\mathbf{q}_{k}^{(n+1)})-R_{k}(\mathbf{q}_{k}^{(n)})<\delta||\mathbf{q}_{k}^{(n+1)}-\mathbf{q}_{k}^{(n)}||^{2}$
$\boldsymbol{\mathbf{\mathrm{or}}}$ $||\mathbf{q}_{k,b_{1}}^{(n+1)}-\mathbf{q}_{k,b_{2}}^{(n+1)}||<d_{\min}$
($\mathrm{for}\:\mathrm{any}\:b_{1}\neq b_{2}$)}{

$\mu_{k}^{(n)}\leftarrow\tau\mu_{k}^{(n)}$\;

}}

}

Obtain $\mathbf{t}^{(n+1)}$ by Algorithm \ref{alg:ALM}\;

$n\leftarrow n+1$ \;

}

\KwOut{$\mathbf{u}^{\mathrm{opt}}$, $\{\mathbf{W}_{k}^{\mathrm{opt}}\}$,
$\mathbf{v}^{\mathrm{opt}}$, $\{\mathbf{q}_{k}^{\mathrm{opt}}\}$}
\end{algorithm}
The overall algorithm for solving the problem (\ref{eq:Orig_WSR_prob}) is summarized in
Algorithm \ref{alg:Overall_alg}. Regarding the convergence of Algorithm
\ref{alg:Overall_alg}, we first observe that the optimization $\mathbf{u}$
does not change the WSR. Moreover, $\{\mathbf{W}_{k}\}$ and $\mathbf{v}$
are optimized by using the concave lower bounds of the objective function
and implementing convex optimization. Therefore, the appropriate objective
values have to satisfy
\begin{align}
&\mathrm{WSR}(\mathbf{u}^{(n+1)},\{\mathbf{W}_{k}^{(n)}\},\mathbf{v}^{(n)},\{\mathbf{q}_{k}^{(n)}\})\nonumber \\
&\qquad\le\mathrm{WSR}(\mathbf{u}^{(n+1)},\{\mathbf{W}_{k}^{(n+1)}\},\mathbf{v}^{(n)},\{\mathbf{q}_{k}^{(n)}\})\nonumber \\
&\qquad\le\mathrm{WSR}(\mathbf{u}^{(n+1)},\{\mathbf{W}_{k}^{(n+1)}\},\mathbf{v}^{(n+1)},\{\mathbf{q}_{k}^{(n)}\}).
\end{align}
{The positions of the users' MAs are optimized independently for each user, 
such that adjusting the MA position of one user affects only its own communication rate, and not the rates of other users.
Therefore, the application of the well-established PGM for the sequential optimization of MA positions for all users leads to an improved objective function value.
Also, the ALM, which optimizes the positions of the BS transmit MAs, increases the objective function.
From the above, we conclude that the objective value is monotonically non-decreasing in each iteration of Algorithm \ref{alg:Overall_alg}. 
Moreover, the objective value is upper-bounded, as demonstrated by the following lemma.
\begin{lem}
The WSR for the system with linear precoding is upper-bounded as \label{lem:WSR_UB}
\begin{align}
\text{WSR}\le\sum_{k=1}^{K}w_{k}N_{u}\ln\left(1+\frac{N_{t}|\rho_{k}|^{2}(P_{\max}+1)}{N_{u}\sigma_{k}^{2}}\right)+w_{k}N_{u}\ln\sigma_{k}^{2}.  \label{eq:WSR_UB_expr}
\end{align}
\end{lem} 
\begin{IEEEproof}
See Appendix C.
\end{IEEEproof}
Hence, Algorithm \ref{alg:Overall_alg} generates a non-decreasing and upper-bounded objective sequence. Since the feasible sets for the used optimization variables are continuous, the objective sequence produced is guaranteed to converge to a fixed point, which is a stationary point of the considered optimization problem. 
}

{The per-iteration complexity of this algorithm is given by
\begin{align}
C_{LP} = &\,\mathcal{O}((KN_{t}N_{u})^{3}+N_{t}^{6}+\nonumber \\
 & K(I_{L,k}+I_{L,T})N_{t}N_{u}(N_{t}+N_{u})),
\end{align}
where $I_{L,k}$ is the number of search steps performed by the line-search procedure when optimizing the MA position of user $k$, and $I_{L,T}$ is the number of search steps performed by the line-search procedure when optimizing the MA positions at the BS.}

\section{Problem Solution for System with ZF Precoding}
Similar to the case of linear precoding, the optimization problem (\ref{eq:Orig_WSR_prob_ZF}) for ZF precoding
 is also non-convex. For this reason, we propose
an AO-based algorithm which individually optimizes each of the optimization variables separately and is elaborated in more detail in the following subsections.

\subsection{Optimization of the Sensing Receive Combiner}
Note that the SINR expression (\ref{eq:SINR_def}) for the case of linear precoding 
is quite similar to the SINR expression (\ref{eq:SINR_def-1}) for the case of ZF precoding.
Therefore, it easy to see that the optimal receive beamformer can be obtained as 
\begin{equation}
\text{\ensuremath{\mathbf{u}}}^{*}=\frac{\mathbf{D}_{z}^{-1}\mathbf{f}_{r}}{||\mathbf{D}_{z}^{-1}\mathbf{f}_{r}||},\label{eq:u_opt-LP}
\end{equation}
where $\mathbf{D}_{z}=\mathbf{G}\mathbf{P}_{e}\mathbf{P}_{e}^{H}\mathbf{G}^{H}+\sigma_{z}^{2}\mathbf{I}_{N_r}$.

\subsection{Optimization of the Sensing Transmit Beamformer}

Similarly as for linear precoding, the sensing transmit beamformer
is optimized using the SCA method. Using the covariance matrix $\mathbf{V}=\mathbf{v}\mathbf{v}^{H}$
and the first-order Taylor approximation
\begin{align}
\ln\left|\mathbf{E}_{k}(\mathbf{V})\right| & \le\ln\left|\mathbf{E}_{k}(\mathbf{V}^{(n)})\right|+\nonumber \\
 & \text{Tr}(\mathbf{H}_{k}^{H}(\mathbf{E}_{k}(\mathbf{V}^{(n)}))^{-1}\mathbf{H}_{k}(\mathbf{V}-\mathbf{V}^{(n)})),
\end{align}
where $\mathbf{E}_{k}(\mathbf{V})=\mathbf{H}_{k}\mathbf{V}\mathbf{H}_{k}^{H}+\sigma_{k}^{2}\mathbf{I}_{N_{u}}$,
$\mathbf{E}_{k}(\mathbf{V}^{(n)})=\mathbf{H}_{k}\mathbf{V}^{(n)}\mathbf{H}_{k}^{H}+\sigma_{k}^{2}\mathbf{I}_{N_{u}}$
and $\mathbf{V}^{(n)}$ denotes $\mathbf{V}$ after $n$ iterations,
the achievable rate of user $k$ can be lower-bounded as
\begin{align}
R_{Z,k}  &\ge\tilde{R}_{Z,k}=\ln\Bigg|\text{\ensuremath{\beta}}_{e}^{2}\mathbf{I}_{N_u}+\mathbf{E}_{k}(\mathbf{V})\Bigg|-\ln\Bigg|\mathbf{E}_{k}(\mathbf{V}^{(n)})\Bigg|\nonumber\\
 & -\text{Tr}(\mathbf{H}_{k}^{H}(\mathbf{E}_{k}(\mathbf{V}^{(n)}))^{-1}\mathbf{H}_{k}(\mathbf{V}-\mathbf{V}^{(n)})).
\end{align}
Following the same steps as in the case of linear precoding, the appropriate
optimization problem can be formulated as
\begin{subequations}
\label{eq:Opt_sens_prec_LP}
\begin{align}
\underset{\mathbf{V}}{\mathrm{maximize}} & \;\sum_{k=1}^{K}w_{k}\tilde{R}_{Z,k}+\zeta\mathcal{M}\\
\text{s.t.} & \;\text{Tr}(\mathbf{V})\le1,\label{eq:Tr_V-2}\\
 & \;(\ref{eq:SINRmin}),\nonumber 
\end{align}
\end{subequations}
 where $\mathcal{M}=\beta_{\max}(\mathbf{V}^{(n)})+\text{Tr}(\boldsymbol{\chi}^{(n)}\boldsymbol{\chi}^{(n)H}(\mathbf{V}-\mathbf{V}^{(n)}))-\text{Tr}(\mathbf{V})$
and $\zeta$ is the penalty parameter. Since this problem is convex,
it can be solved by any conventional optimization solver, and the
appropriate algorithm has the same form as  Algorithm~\ref{alg:Prec_sense_vector}.

\subsection{Optimization of MA Positions of User $k$}
Changing the MA positions of a single user in a system with ZF precoding
affects the achievable rate of all users, as well as the sensing SINR.
This is a major difference compared to systems with linear precoding,
where any change the MA positions of a user influences only its the
achievable rate. Therefore, the optimal MA positions of user $k$
in the considered system with ZF preceding are determined by the following
optimization problem: 
\begin{subequations}
\label{eq:Opt_qk-1}
\begin{align}
\underset{\{\mathbf{q}_{k,b} \}}{\mathrm{maximize}} & \;\text{WSR}=\sum_{k=1}^{K}w_{k}R_{Z,k}\\
\text{s.t.} & \;(\ref{eq:SINRmin}),(\ref{eq:user_region}),(\ref{eq:User_space}), {(\ref{eq:rank_cons})}.\label{eq:Tr_V-1-1}
\end{align}
\end{subequations}

For convenience, the SINR constraint (\ref{eq:SINRmin}) for ZF precoding
can be reformulated as
\begin{gather}
\kappa_{z}\triangleq\mathbf{u}^{H}\mathbf{G}(\gamma_{0}\mathbf{P}_{e}\mathbf{P}_{e}^{H}-\mathbf{v}\mathbf{v}^{H})\mathbf{G}^{H}\mathbf{u}+\gamma_{0}\sigma_{z}^{2}\mathbf{u}^{H}\mathbf{u}\le0.\label{eq:SINR_ref-ZF}
\end{gather}

To deal with this, similarly as in subsection \ref{subsec:ZF-Prec-Opt-BS-TX},
we use the ALM and reformulate (\ref{eq:Opt_qk-1})
according to 
\begin{subequations}
\label{eq:Lagr-1}
\begin{align}
\text{\ensuremath{\underset{\{\mathbf{q}_{k,b} \}}{\mathrm{minimize}}}} & \;L(\mathbf{q}_{k})=-\sum_{k=1}^{K}w_{k}R_{Z,k}+\eta\kappa_{z}+\frac{1}{2}p\kappa_{z}^{2}\label{eq:Lt_obj-1}\\
\text{s.t.} & \;(\ref{eq:user_region}),(\ref{eq:User_space}), {(\ref{eq:rank_cons})},\label{eq:SINRmin-1-2-1}
\end{align}
\end{subequations}
 where $\eta$ is the Lagrangian multiplier and the penalty factor
is given by 
\begin{equation}
p=\begin{cases}
0 & \kappa_{z}<0
\:\text{and}\:\eta=0,
\\
p_{0} & \text{otherwise}.
\end{cases}
\end{equation}
We employ the PGM to solve (\ref{eq:Lagr-1}) via
\begin{equation}
\mathbf{q}_{k}^{(n+1)}=P(\mathbf{q}_{k}^{(n)}-\alpha^{(n)}\nabla_{\mathbf{q}_{k}}L(\mathbf{q}_{k}^{(n)})),
\end{equation}
where $\alpha^{(n)}$ is the step size. The gradient $\nabla_{\mathbf{q}_{k}}L(\mathbf{q}_{k}^{(n)})$
is determined by the gradients of $L(\mathbf{q}_{k}^{(n)})$ w.r.t.
the coordinates of individual MAs of user $k$. For the $b$-th ($b\in\{1,\dots,N_{u}\}$)
MA of user $k$, these gradients are given by 
\begin{align}
\nabla_{x_{k,b}}L(\mathbf{q}_{k}^{(n)})\!=\!-\sum_{k=1}^{K}w_{k}\nabla_{x_{k,b}}R_{Z,u}+(\eta+p\kappa_z)\nabla_{x_{k,b}}\kappa_z,\\
\nabla_{y_{k,b}}L(\mathbf{q}_{k}^{(n)})\!=\!-\sum_{k=1}^{K}w_{k}\nabla_{y_{k,b}}R_{Z,u}+(\eta+p\kappa_z)\nabla_{y_{k,b}}\kappa_z,
\end{align}
where expressions for the gradients of $R_{Z,u}$ and $\kappa_{z}$ w.r.t. the specified
MA coordinates are given in the following lemmas.
\begin{lem}
The gradients of $R_{Z,u}$ ($u\in\{1,\dots,K\}$) w.r.t. the coordinates
of the b-th MA of user k are given by (\ref{eq:dRu_xkb}) and (\ref{eq:dRu_ykb}) on the next page,
where $\delta_{k,u}$ is the Kronecker delta defined as 
\begin{figure*}[tbh]
\begin{align}
\nabla_{x_{k,b}}R_{Z,u} & =-\frac{4\pi\rho_{k}}{\lambda}\mathfrak{I}\Big(\sum_{m=1}^{N_{t}}\left[P_{\max}T^{-2}\mathrm{Tr}(\mathbf{F}_{u}^{-1})\mathbf{B}_{e}(m,b_{r})+\delta_{k,u}\mathbf{G}_{k}(m,b)\right]e^{j\frac{2\pi}{\lambda}||\mathbf{t}_{m}-\mathbf{q}_{k,b}||}\frac{x_{k,b}-x_{m}}{||\mathbf{t}_{m}-\mathbf{q}_{k,b}||}\Big)\label{eq:dRu_xkb}\\
\nabla_{y_{k,b}}R_{Z,u} & =-\frac{4\pi\rho_{k}}{\lambda}\mathfrak{I}\Big(\sum_{m=1}^{N_{t}}\left[P_{\max}T^{-2}\mathrm{Tr}(\mathbf{F}_{u}^{-1})\mathbf{B}_{e}(m,b_{r})+\delta_{k,u}\mathbf{G}_{k}(m,b)\right]e^{j\frac{2\pi}{\lambda}||\mathbf{t}_{m}-\mathbf{q}_{k,b}||}\frac{y_{k,b}-y_{m}}{||\mathbf{t}_{m}-\mathbf{q}_{k,b}||}\Big)\label{eq:dRu_ykb}
\end{align}
\hrule
\vspace{-1.5em}
\end{figure*}
\[
\delta_{k,u}=\begin{cases}
1 & k=u,\\
0 & k\neq u,
\end{cases}
\]
$b_{k}=(k-1)N_{u}+b$, $T=\mathrm{Tr}((\mathbf{H}_{e}\mathbf{H}_{e}^{H})^{-1})$,
$\mathbf{B}_{e}=\mathbf{H}_{e}^{H}(\mathbf{H}_{e}\mathbf{H}_{e}^{H})^{-2},$
$\mathbf{E}_{k}=\mathbf{H}_{k}\mathbf{v}\mathbf{v}^{H}\mathbf{H}_{k}^{H}+\sigma_{k}^{2}\mathbf{I}_{N_{u}}$,
$\mathbf{F}_{k}=\text{\ensuremath{\beta_{e}^{2}}}\mathbf{I}_{N_u}+\mathbf{H}_{k}\mathbf{v}\mathbf{v}^{H}\mathbf{H}_{k}^{H}+\sigma_{k}^{2}\mathbf{I}_{N_{u}}$,
$\mathbf{F}_{u}=\text{\ensuremath{\beta_{e}^{2}}}\mathbf{I}_{N_{u}}+\mathbf{H}_{u}\mathbf{v}\mathbf{v}^{H}\mathbf{H}_{u}^{H}+\sigma_{k}^{2}\mathbf{I}_{N_{u}}$
and $\mathbf{G}_{k}=\mathbf{v}\mathbf{v}^{H}\mathbf{H}_{k}^{H}(\mathbf{F}_{k}^{-1}-\mathbf{E}_{k}^{-1})$.
\label{lem:dRk_UserMAPos}
\end{lem}
\begin{IEEEproof}
See Appendix D.
\end{IEEEproof}
\begin{lem}
The gradients of $\kappa_{z}$ w.r.t. the coordinates of the b-th
MA of user k are given by (\ref{eq:dSINR_xkb}) and (\ref{eq:dSINR_ykb}) on the next page,
where $b_{k}=(k-1)N_{u}+b$, $T=\mathrm{Tr}((\mathbf{H}_{e}\mathbf{H}_{e}^{H})^{-1})$,
$\mathbf{B}_{e}=\mathbf{H}_{e}^{H}(\mathbf{H}_{e}\mathbf{H}_{e}^{H})^{-2}$,
$\mathbf{P}_{1}=\mathbf{H}_{e}^{H}(\mathbf{H}_{e}\mathbf{H}_{e}^{H})^{-1}$,
$\mathbf{D}=\mathbf{P}_{1}^{H}\mathbf{G}^{H}\text{\ensuremath{\mathbf{u}}}\text{\ensuremath{\mathbf{u}}}^{H}\mathbf{G}$,
$\mathbf{D}_{1}=\mathbf{H}_{e}^{H}\mathbf{D}\mathbf{H}_{e}^{H}(\mathbf{H}_{e}\mathbf{H}_{e}^{H})^{-2}$
and $\mathbf{D}_{2}=(\mathbf{H}_{e}\mathbf{H}_{e}^{H})^{-1}\mathbf{D}-\mathbf{D}\mathbf{H}_{e}^{H}(\mathbf{H}_{e}\mathbf{H}_{e}^{H})^{-2}\mathbf{H}_{e}.$
\label{lem:dSINR_UserMAPos}
\begin{figure*}[tbh]
\begin{align}
\nabla_{x_{k,b}}\kappa_{z} & =-\gamma_{0}\text{\ensuremath{\mathbf{u}}}^{H}\mathbf{G}\mathbf{P}_{1}\mathbf{P}_{1}^{H}\mathbf{G}^{H}\text{\ensuremath{\mathbf{u}\frac{4\pi\rho_{k}}{\lambda}P_{\max}T^{-2}\mathfrak{I}\left(\sum_{m=1}^{N_{t}}\mathbf{B}_{e}(m,b_{r})e^{j\frac{2\pi}{\lambda}||\mathbf{t}_{m}-\mathbf{q}_{k,b}||}\frac{x_{k,b}-x_{m}}{||\mathbf{t}_{m}-\mathbf{q}_{k,b}||}\right)}}\nonumber \\
 & +\gamma_{0}\beta_{e}^{2}\frac{4\pi\rho_{k}}{\lambda}\mathfrak{I}\Big(\sum_{m=1}^{N_{t}}\left[\mathbf{D}_{1}(m,b_{r})e^{j\frac{2\pi}{\lambda}||\mathbf{t}_{m}-\mathbf{q}_{k,b}||}+\mathbf{D}_{2}^{T}(m,b_{r})e^{-j\frac{2\pi}{\lambda}||\mathbf{t}_{m}-\mathbf{q}_{k,b}||}\right]\frac{x_{k,b}-x_{m}}{||\mathbf{t}_{m}-\mathbf{q}_{k,b}||}\Big)\label{eq:dSINR_xkb}
\end{align}
\begin{align}
\nabla_{y_{k,b}}\kappa_{z} & =-\gamma_{0}\text{\ensuremath{\mathbf{u}}}^{H}\mathbf{G}\mathbf{P}_{1}\mathbf{P}_{1}^{H}\mathbf{G}^{H}\text{\ensuremath{\mathbf{u}\frac{4\pi\rho_{k}}{\lambda}P_{\max}T^{-2}\mathfrak{I}\Big(\sum_{m=1}^{N_{t}}\mathbf{B}_{e}(m,b_{r})e^{j\frac{2\pi}{\lambda}||\mathbf{t}_{m}-\mathbf{q}_{k,b}||}\frac{y_{k,b}-y_{m}}{||\mathbf{t}_{m}-\mathbf{q}_{k,b}||}\Big)}}\nonumber \\
 & +\gamma_{0}\beta_{e}^{2}\frac{4\pi\rho_{k}}{\lambda}\mathfrak{I}\Big(\sum_{m=1}^{N_{t}}\left[\mathbf{D}_{1}(m,b_{r})e^{j\frac{2\pi}{\lambda}||\mathbf{t}_{m}-\mathbf{q}_{k,b}||}+\mathbf{D}_{2}^{T}(m,b_{r})e^{-j\frac{2\pi}{\lambda}||\mathbf{t}_{m}-\mathbf{q}_{k,b}||}\right]\frac{y_{k,b}-y_{m}}{||\mathbf{t}_{m}-\mathbf{q}_{k,b}||}\Big)\label{eq:dSINR_ykb}
\end{align}
\hrule
\vspace{-1.5em}
\end{figure*}
\end{lem}
\begin{IEEEproof}
See Appendix E.
\end{IEEEproof}
The algorithm for optimizing the MA positions of user $k$ is outlined
in Algorithm \ref{alg:ALM-qk} and is very similar to Algorithm \ref{alg:ALM}. {A small difference is that for ZF precoding, the backtracking line search needs to take into account the constraint  (\ref{eq:rank_cons}), as shown in Algorithm \ref{alg:ALM-qk}}. 
One should note that, due to ZF precoding, any change in the coordinates
of the MAs of user $k$ requires re-computation of the appropriate channel
matrix $\mathbf{H}_{k}$ as well as the precoding matrix $\mathbf{P}_{e}$.
On the other hand, only $\mathbf{H}_{k}$ needs to be recomputed in
the case of linear precoding.

\begin{algorithm}[t]
\small
\caption{ALM for Optimizing the MA Positions of User $k$\label{alg:ALM-qk}}

\DontPrintSemicolon
\LinesNumbered 

\KwIn{$\mathbf{q}_{k}^{(0)}$, $\eta\leftarrow0$, $p_{0}$, $\theta$}

$i\leftarrow0$\;

\Repeat{ $|\mathrm{WSR}(\mathbf{q}_{\text{c}}^{(i)})-\mathrm{WSR}(\mathbf{q}_{\text{c}}^{(i-1)})|<\epsilon_{f}$}{

Calculate $L(\mathbf{q}_{k}^{(0)})$ according to (\ref{eq:Lt_obj-1})

$n\leftarrow0$\;
\Repeat{$\left|\frac{L(\mathbf{q}_{k}^{(n)})-L(\mathbf{q}_{k}^{(n-1)})}{L(\mathbf{q}_{k}^{(n)})}\right|<\epsilon_{L}$}{

\Repeat{$L(\mathbf{q}_{k}^{(n)})-L(\mathbf{q}_{k}^{(n+1)})\ge\delta||\mathbf{q}_{k}^{(n+1)}-\mathbf{q}_{k}^{(n)}||^{2}$
$\mathrm{\mathbf{and}}$ $||\mathbf{q}_{k,b_{1}}^{(n+1)}-\mathbf{q}_{k,b_{2}}^{(n+1)}||\ge d_{\min}\:\:(\mathrm{for}\:\mathrm{all}\:b_{1}\neq b_{2})$
$\mathrm{\mathbf{and}}$
{$\mathrm{rank}(\mathbf{H}_{e})=KN_{u}$}}{

$\mathbf{q}_{k}^{(n+1)}=P(\mathbf{q}_{k}^{(n)}-\alpha^{(n)}\nabla_{\mathbf{q}_{k}}L(\mathbf{q}_{k}^{(n)}))$\;

\If{$L(\mathbf{q}_{k}^{(n)})-L(\mathbf{q}_{k}^{(n+1)})<\delta||\mathbf{q}_{k}^{(n+1)}-\mathbf{q}_{k}^{(n)}||^{2}$
$\boldsymbol{\mathbf{\mathrm{or}}}$ $||\mathbf{q}_{k,b_{1}}^{(n+1)}-\mathbf{q}_{k,b_{2}}^{(n+1)}||<d_{\min}\:\:(\mathrm{for}\:\mathrm{any}\:b_{1}\neq b_{2})$ $\boldsymbol{\mathbf{\mathrm{or}}}$
{$\mathrm{rank}(\mathbf{H}_{e})<KN_{u}$}
} {

$\alpha^{(n)}\leftarrow\tau\alpha^{(n)}$\;

}}

$n\leftarrow n+1$\;

}

$\mathbf{q}_{\text{c}}^{(i)}=\mathbf{q}_{k}^{(n)}$\;

$\mathbf{q}_{k}^{(0)}=\mathbf{q}_{k}^{(n)}$\;

$\eta=\max(0,\eta+p_{0}\kappa_{z})$\;

$p_{0}=\theta p_{0}$\;

$i\leftarrow i+1$

}

\KwOut{$\mathbf{q}_{k}^{*}=\mathbf{q}_{\text{c}}^{(i)}$}
\end{algorithm}

\subsection{Optimization of the BS Transmit MA Positions}

The optimal BS transmit MA positions are determined by solving the following problem:
\begin{subequations}
\label{eq:Opt_qk-1-1}
\begin{align}
\underset{\{\mathbf{t}_m\}}{\mathrm{maximize}} & \;\text{WSR}=\sum_{k=1}^{K}w_{k}R_{Z,k}\\
\text{s.t.} & \;(\ref{eq:BS_region}),(\ref{eq:TX_space}),{ (\ref{eq:rank_cons})}, (\ref{eq:SINR_ref-ZF}).\label{eq:Tr_V-1-1-1}
\end{align}
\end{subequations}

This problem can be solved using same optimization method and the
same steps as in the previous subsection. The only difference is in
the gradients of the WSR and the sensing SINR $\kappa_{z}$ w.r.t.
the transmit MA coordinates, which are specified in the following
lemmas.
\begin{lem}
The gradients of $R_{Z,u}$ ($u\in\{1,\dots,K\}$) w.r.t. the coordinates
of the m-th BS transmit MA are given by (\ref{eq:dRu_xm}) and (\ref{eq:dRu-ym}) on the next page,
where 
$\mathbf{F}_{u}=\text{\ensuremath{\beta_{e}^{2}}}\mathbf{I}_{N_u}+\mathbf{H}_{u}\mathbf{v}\mathbf{v}^{H}\mathbf{H}_{u}^{H}+\sigma_{k}^{2}\mathbf{I}_{N_{u}}$
and $\mathbf{G}_{u}=\mathbf{v}\mathbf{v}^{H}\mathbf{H}_{u}^{H}(\mathbf{F}_{u}^{-1}-\mathbf{E}_{u}^{-1})$.
\begin{figure*}[tbh]
\begin{align}
\nabla_{x_{m}}R_{Z,u}= & -\frac{4\pi}{\lambda}\mathfrak{I}\Big(P_{\max}T^{-2}\text{Tr}(\mathbf{F}_{u}^{-1})\sum_{k=1}^{K}\sum_{b=1}^{N_{u}}\rho_{k}\mathbf{B}_{e}(m,b_{r})e^{j\frac{2\pi}{\lambda}||\mathbf{t}_{m}-\mathbf{q}_{k,b}||}\frac{x_{m}-x_{k,b}}{||\mathbf{t}_{m}-\mathbf{q}_{k,b}||}\Big)\nonumber \\
 & -\frac{4\pi\rho_{u}}{\lambda}\mathfrak{I}\Big(\sum_{b=1}^{N_{u}}\mathbf{G}_{u}(m,b)e^{j\frac{2\pi}{\lambda}||\mathbf{t}_{m}-\mathbf{q}_{u,b}||}\frac{x_{m}-x_{u,b}}{||\mathbf{t}_{m}-\mathbf{q}_{u,b}||}\Big),\label{eq:dRu_xm}\\
\nabla_{y_{m}}R_{Z,u}= & -\frac{4\pi}{\lambda}\mathfrak{I}\Big(P_{\max}T^{-2}\text{Tr}(\mathbf{F}_{u}^{-1})\sum_{k=1}^{K}\sum_{b=1}^{N_{u}}\rho_{k}\mathbf{B}_{e}(m,b_{r})e^{j\frac{2\pi}{\lambda}||\mathbf{t}_{m}-\mathbf{q}_{k,b}||}\frac{y_{m}-y_{k,b}}{||\mathbf{t}_{m}-\mathbf{q}_{k,b}||}\Big)\nonumber \\
 & -\frac{4\pi\rho_{u}}{\lambda}\mathfrak{I}\Big(\sum_{b=1}^{N_{u}}\mathbf{G}_{u}(m,b)e^{j\frac{2\pi}{\lambda}||\mathbf{t}_{m}-\mathbf{q}_{u,b}||}\frac{y_{m}-y_{u,b}}{||\mathbf{t}_{m}-\mathbf{q}_{u,b}||}\Big)\label{eq:dRu-ym}
\end{align}
\hrule
\end{figure*}
\end{lem}
\begin{IEEEproof}
The proof is similar to the proof of Lemma \ref{lem:dRk_UserMAPos}.
\end{IEEEproof}
\begin{lem}
The gradients of $\kappa_{z}$ w.r.t. the coordinates of the m-th
BS transmit MA are given by (\ref{eq:dSINR_xm}) and (\ref{eq:dSINR_ym}) on the next page.
\begin{figure*}[tbh]
\begin{align}
\nabla_{x_{m}}\kappa_{z} & =-\frac{4\pi\rho_{s}}{\lambda}\mathfrak{I}\Big(\mathbf{\mathbf{e}}(m)e^{j\frac{2\pi}{\lambda}||\mathbf{t}_{m}-\mathbf{s}_{l}||}\frac{x_{m}-x_{s}}{||\mathbf{t}_{m}-\mathbf{s}||}\Big)\nonumber \\
 & -\frac{4\pi}{\lambda}P_{\max}T^{-2}\gamma_{0}\text{\ensuremath{\mathbf{u}}}^{H}\mathbf{G}\mathbf{P}_{1}\mathbf{P}_{1}^{H}\mathbf{G}^{H}\mathbf{u}\mathfrak{I}\text{\ensuremath{\Big(\sum_{k=1}^{K}\sum_{b=1}^{N_{u}}\rho_{k}\mathbf{B}_{e}(m,b_{r})e^{j\frac{2\pi}{\lambda}||\mathbf{t}_{m}-\mathbf{q}_{k,b}||}\frac{x_{m}-x_{k,b}}{||\mathbf{t}_{m}-\mathbf{q}_{k,b}||}\Big)}}\nonumber \\
 & +\frac{4\pi}{\lambda}\gamma_{0}\beta^{2}\mathfrak{I}\Big(\sum_{k=1}^{K}\sum_{b=1}^{N_{u}}\rho_{k}\left[\mathbf{D}_{1}(m,b_{r})e^{j\frac{2\pi}{\lambda}||\mathbf{t}_{m}-\mathbf{q}_{k,b}||}+\mathbf{D}_{2}^{T}(m,b_{r})e^{-j\frac{2\pi}{\lambda}||\mathbf{t}_{m}-\mathbf{q}_{k,b}||}\right]\frac{x_{m}-x_{k,b}}{||\mathbf{t}_{m}-\mathbf{q}_{k,b}||}\Big)\label{eq:dSINR_xm}
\end{align}
\begin{align}
\nabla_{y_{m}}\kappa_{z} & =-\frac{4\pi\rho_{s}}{\lambda}\mathfrak{I}\Big(\mathbf{\mathbf{e}}(m)e^{j\frac{2\pi}{\lambda}||\mathbf{t}_{m}-\mathbf{s}_{l}||}\frac{x_{m}-x_{s}}{||\mathbf{t}_{m}-\mathbf{s}||}\Big)\nonumber \\
 & -\frac{4\pi}{\lambda}P_{\max}T^{-2}\gamma_{0}\text{\ensuremath{\mathbf{u}}}^{H}\mathbf{G}\mathbf{P}_{1}\mathbf{P}_{1}^{H}\mathbf{G}^{H}\text{\ensuremath{\mathbf{u}\mathfrak{I}\Big(\sum_{k=1}^{K}\sum_{b=1}^{N_{u}}\rho_{k}\mathbf{B}_{e}(m,b_{r})e^{j\frac{2\pi}{\lambda}||\mathbf{t}_{m}-\mathbf{q}_{k,b}||}\frac{y_{m}-y_{k,b}}{||\mathbf{t}_{m}-\mathbf{q}_{k,b}||}\Big)}}\nonumber \\
 & +\frac{4\pi}{\lambda}\gamma_{0}\beta^{2}\mathfrak{I}\Big(\sum_{k=1}^{K}\sum_{b=1}^{N_{u}}\rho_{k}\left[\mathbf{D}_{1}(m,b_{r})e^{j\frac{2\pi}{\lambda}||\mathbf{t}_{m}-\mathbf{q}_{k,b}||}+\mathbf{D}_{2}^{T}(m,b_{r})e^{-j\frac{2\pi}{\lambda}||\mathbf{t}_{m}-\mathbf{q}_{k,b}||}\right]\frac{y_{m}-y_{k,b}}{||\mathbf{t}_{m}-\mathbf{q}_{k,b}||}\Big)\label{eq:dSINR_ym}
\end{align}
\hrule%
\end{figure*}
\end{lem}

\begin{IEEEproof}
The proof is similar to the proof of Lemma \ref{lem:dSINR_UserMAPos}.
\end{IEEEproof}

\subsection{Overall Algorithm}
Due to the similarity with Algorithm \ref{alg:Overall_alg},
we provide only a description of the overall algorithm. In each iteration,
the sensing receive combiner is obtained by using (\ref{eq:u_opt-LP}),
which provides a closed-form solution. The sensing transmit beamformer
is optimized by the SCA method for which we use a concave lower bound
on the user's achievable rate; therefore, it improves the WSR of
the system. Also, the MA positions are optimized separately for each
user with the ALM, presented in Algorithm \ref{alg:ALM-qk}, which
provably increases the WSR. Using the same optimization method, the
positions of the BS transmit MAs are adjusted.
{Hence, we can conclude that the objective value is monotonically non-decreasing in each iteration of this algorithm. Similarly to Lemma \ref{lem:WSR_UB}, it can be shown that the objective is upper-bounded. Since the feasible sets for the used optimization variables are continuous, the objective sequence is guaranteed to converge to a fixed point, which is a stationary point of the considered optimization problem.}

{The complexity of this algorithm is given by
\begin{align}
C_{ZF} = & \,\mathcal{O}(N_{t}^{6}+I_{Z,k}K[(KN_{u})^{3}+N_{t}(KN_{u})^{2}+N_{t}^{2}N_{u}]\nonumber \\
 & +I_{Z,T}[(KN_{u})^{3}+N_{t}(KN_{u})^{2}+KN_{t}^{2}N_{u}]),
\end{align}
where $I_{Z,k}$ is the number of search steps performed by the line-search procedure when optimizing the MA position of user $k$, and $I_{Z,T}$ is the number of search steps performed by the line-search procedure when optimizing the MA positions at the BS.}

\section{Simulation Results}
In this section, we evaluate the WSR of the considered schemes with
linear precoding (denoted as LP-MA) and ZF precoding (denoted as ZF-MA)
by means of Monte Carlo simulations. As benchmarks, we consider schemes that differ from the proposed ones in that the BS transmitter and users have fixed antennas; these are referred to as the LP-FIX and ZF-FIX scheme, respectively. 
The initial positions of the
users' MAs and the BS transmit MAs are the same for all four 
schemes and are randomly chosen inside the MA regions ${C}_{k}$
and ${C}_{t}$, respectively. The free space path loss between
the BS and user $k$ is modeled as $\rho_{k}=\lambda^{2}/(4\pi d_{tk})^{2}$
\cite{liu2023near}, where $d_{tk}=\|\mathbf{o}_{t}-\mathbf{o}_{k}\|$
is the distance between the BS transmit MA array and user $k$. The round-trip
channel coefficient for target sensing is given by $\rho_{s}=\lambda^{2}/((4\pi)^{3}R_{t}^{2}R_{r}^{2})$
\cite{dong2022sensing}, where $R_{t}=\|\mathbf{o}_{t}-\mathbf{s}\|$
is the distance between the BS transmit MA array and the target, and
$R_{r}=\|\mathbf{o}_{r}-\mathbf{s}\|$ is the distance between the BS
receive antenna array and the target.

In the following simulation setup, the parameters are $f=30\,\mathrm{GHz}$
(i.e., $\lambda=1\,\mathrm{cm}$), $N_{t}=16$, $N_{r}=8$, $K=2$,
$N_{u}=4$, $P_{\max}=1\,\mathrm{W}$, $d_{\min}=\lambda/2=0.5\,\mathrm{cm}$,
$\gamma_{0}=0.01=-20\,\mathrm{dB}$, $\zeta=1$ and $\sigma_{k}^{2}=\sigma_{z}^{2}=-100\,\mathrm{dB}$.
The center of the BS transmit MA region is located $\mathbf{o}_{t}=[-3\,\mathrm{m},10\,\mathrm{m},0]^{T}$
and the length of its side is $L_{t}=1\,\mathrm{m}$. The midpoint
of the receive BS ULAs is located at $\mathbf{o}_{r}=[3\,\mathrm{m},10\,\mathrm{m},0]^{T}$
and the length of this ULA is $L_{r}=1\,\mathrm{m}$. Both users
are placed at the same distance from the BS transmit MA array in the
directions of $\pm\pi/4$ in the $xz$-plane. The centers of the users'
MA regions are located at $\mathbf{o}_{u,1}=[-3\,\mathrm{m}+d_{0}\sin(\pi/4),1.5\,\mathrm{m},d_{0}\cos(\pi/4)]^{T}$
and $\mathbf{o}_{u,2}=[-3\,\mathrm{m}+d_{0}\sin(-\pi/4),1.5\,\mathrm{m},d_{0}\cos(-\pi/4)]^{T}$,
where $d_{0}=30\,\mathrm{m}$, and the side length of each of these
regions is $a_{k}=15\,\mathrm{cm}$. The initial positions of the
users' antennas are randomly selected inside the specified MA regions.
The point target is located at $\mathbf{s}=[10\,\mathrm{m},1.5\,\mathrm{m},10\,\mathrm{m}]^{T}$.
The CVX tool is used to solve (\ref{eq:Prec_opt_prob}), (\ref{eq:Opt_sense_prec_vec}) and (\ref{eq:Opt_sens_prec_LP}).
In the line search procedure for the PGM, all step sizes are initially
set to $1$, $\delta=10^{-2}$ and $\tau=1/2$. The parameters of
the ALM are initialized as $\eta=0$, $p_{0}=1$,
$\theta=10$, $\epsilon_{L}=10^{-3}$ and $\epsilon_{f}=10^{-2}$.
For the SCA method, the convergence parameter is $\epsilon_{s}=10^{-2}$.
All results are averaged over 100 independent channel realizations.

\begin{figure}[t]
\centering{}\includegraphics[scale=0.85]{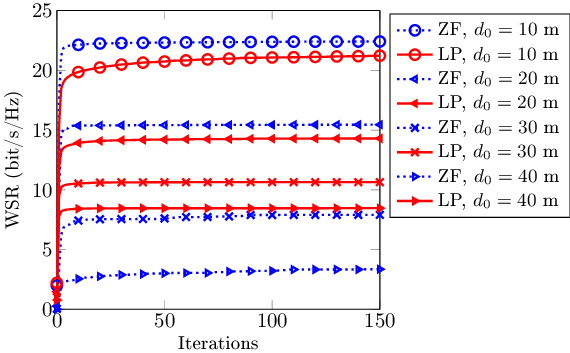}\caption{Convergence of the proposed algorithms. \label{fig:Convergence}}
\end{figure}

The convergence of the proposed algorithms for different values of the distance between the BS transmit MA array and users on the ground \textit{xz}-plane, $d_{0}$, is shown in Fig. \ref{fig:Convergence}.
In general, both of the proposed schemes require a relatively low number of iterations
to converge. When the users are up to 20~m away from the BS transmit
MAs, the ZF-MA scheme achieves a larger WSR than the LP-MA scheme. Since the considered users' channels are modeled similarly as pure \ac{LOS}
channels and users are not placed close to each other, the influence
of multi-user interference very limited. Additionally, the inter-antenna interference between the receive MAs of a single user\footnote{{This interference is determined by the correlation between the responses of different MAs of a single user. If these responses (i.e., rows of the appropriate channel matrix) are denoted by $\mathbf{h}_{i}$ and $\mathbf{h}_{j}$, then their mutual correlation is given by ${\mathbf{h}_{i}\cdot\mathbf{h}_{j}}/{\left\Vert \mathbf{h}_{i}\right\Vert \left\Vert \mathbf{h}_{j}\right\Vert}$, where $\cdot$ presents dot product.}
} 
is relatively weak for users placed in the vicinity of the BS. Therefore, we can conclude that for MA-enabled ISAC systems operating in the near field, ZF precoding can achieve a larger WSR in the interference limited case.
On the other hand, for larger $d_{0}$,
the LP-MA scheme provides better WSR performance. In this scenario, the inter-antenna interference between the receive MAs of a single user is increased and
the appropriate channel matrices are ill-conditioned
(i.e., the corresponding condition numbers are large). Hence, the LP-MA
scheme performs an unequal power allocation among different users'
precoding matrices, as well as among different sub-channels of the same user, which is the optimal WSR achieving strategy in this case
(i.e., waterfilling power allocation). In contrast, the ZF-MA scheme  
forms a transmission system of orthogonal sub-channels with the equal
power gain (i.e., $\mathbf{H}_{e}\mathbf{P}_{e}=\beta_{e}\mathbf{I}_{KN_u}$),
resulting in worse system performance for larger $d_{0}$.

\begin{figure}[t]
\centering{}\includegraphics[scale=0.85]{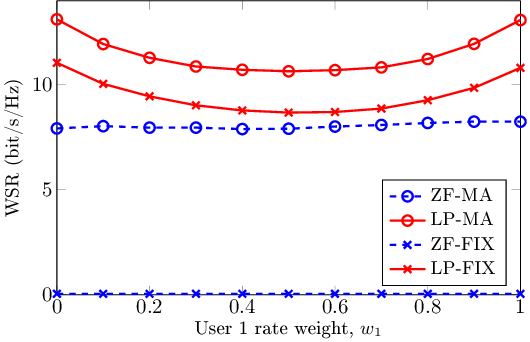}\caption{{WSR versus the rate weight of user 1.} \label{fig:WSR-vs-weight}}
\end{figure}

\begin{figure}[t]
\centering{}\includegraphics[scale=0.85]{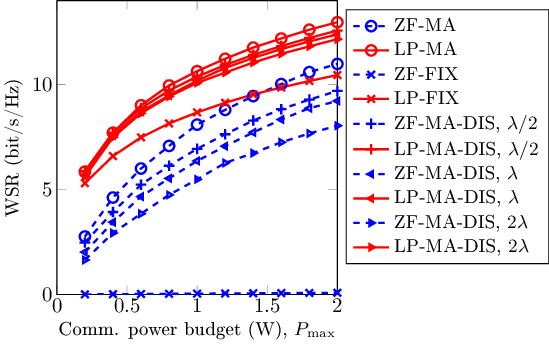}\caption{WSR versus $P_{\max}$.\label{fig:WSR-vs-power}\label{fig:Convergence}}
\end{figure}

The WSR for different values of the users’ rate weights is presented
in Fig. \ref{fig:WSR-vs-weight}, where we have imposed the condition
$w_{1}+w_{2}=1$. The  ZF-MA scheme 
achieves almost constant WSR irrespective of the weighted rate contributions
of both users. This is most likely caused by the fact that ZF precoding
creates a system of orthogonal sub-channels with equal power
gain, and consequently approximately equal achievable rate for every sub-channel.
Since both users have the same number of antennas (i.e., sub-channels), any change
of the users' weight rates has a negligible influence on the
WSR. On the other hand, the WSR for the ZF-MA scheme 
is affected to a significantly greater degree by the users' weight rates. The largest WSR is achieved when most of data transmission is allocated to a single
user, while the lowest WSR is obtained when the users' weighted rate contributions are approximately equal. This suggests that linear precoding is particularly beneficial in systems with a significant imbalance between the users' rate weights. Regarding the schemes with only fixed antennas, we observe that the LP-FIX scheme achieves around 2 bit/s/Hz lower WSR compared to the LP-MA scheme, while the WSR for the ZF-MA scheme remains negligible for all users' rate weights.

Next, we study how the WSR varies with the {communication power budget}
(i.e., $P_{\max}$), as shown in Fig. \ref{fig:WSR-vs-power}. { For a comprehensive performance evaluation, we additionally consider two benchmark schemes with discrete MA-position sets, denoted as LP-MA-DIS and ZF-MA-DIS. The minimum distances between discrete MA-positions for these two schemes are $\lambda/2$, $\lambda$, and $2\lambda$.}   
The
WSR curves for both proposed schemes exhibit an approximately logarithmic shape
due to the logarithmic increase in the users’ achievable rates. Moreover,
the performance gap between these two schemes tends to gradually decrease
with $P_{\max}$. It seems that the ability of linear precoding to adaptively
allocate power offers less WSR advantage for {a larger communication power budget}.
A similar effect occurs in conventional point-to-point MIMO systems,
where the uniform power allocation becomes optimal at high power levels.
The WSR for the LP-FIX scheme is lower than the WSR for the LP-MA scheme and this difference gradually increases with $P_{\max}$. Similarly, as in the previous figure, the WSR for the ZF-FIX scheme stays negligibly small irrespective of {the communication power budget}.{     
Regarding the schemes with discrete MA-position sets, we observe that the WSRs for both schemes are decreased with the minimum inter-MA distances. Moreover, this reduction is much more pronounced for the ZF-MA-DIS scheme than for the LP-MA-DIS scheme. }

\begin{figure}[t]
\centering{}\includegraphics[scale=0.85]{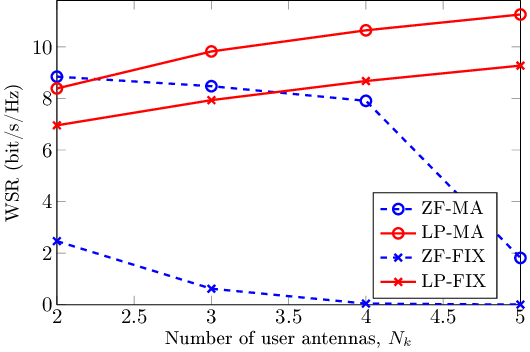}\caption{WSR versus $N_u$.\label{fig:WSR-vs-Nk}}
\end{figure}

In Fig. \ref{fig:WSR-vs-Nk}, we present the WSR for different numbers
of MAs at each user. For low $N_{u}$, when the inter-antenna interference between different MAs of a  single user is relatively small, the ZF-MA scheme provides the best WSR performance. Interestingly, even the ZF-FIX scheme is capable of providing a moderate WSR, albeit lower than other schemes. However, for larger $N_{u}$, the aforementioned interference is increased, leading to a reduction in the WSR for these two schemes. This reduction can be quite severe, as can be seen for the WSR for the ZF-MA scheme when $N_{u}$
changes from 4 to 5. 
In contrast to this, the WSRs for the LP-MA and the LP-FIX scheme steadily increase with $N_{u}$, due to the ability of linear precoding to adaptively allocate power.  
This suggests that linear precoding can be particularly useful for future communication systems, where users' terminals are expected to have a large number of antennas.

\begin{figure}[t!]
\centering{}\subfloat[{Users' MA regions.}]{\centering{}\includegraphics{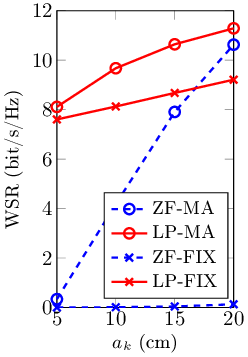}}\subfloat[{BS MA region.}]{\centering{}\includegraphics{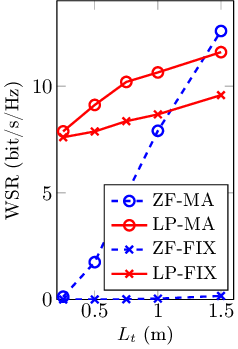}}\caption{{WSR versus the sizes of the users' and
BS MA regions.} \label{fig:WSR-vs-regions}}
\end{figure}

{In Fig. \ref{fig:WSR-vs-regions}, we present the WSR for different sizes of the users' and BS MA regions (i.e., $a_k$ and $L_t$). We notice that the WSR for ZF-MA scheme increases significantly faster with the size of the users' MA regions $a_k$ compared to the WSR for the ZF-LP scheme. The reason is that ZF precoding is more sensitive to the correlation between the impulse responses of users' MAs, as already shown in Fig. 5. A similar trend can be seen with the change of the size of the BS MA region $L_t$, and in this case the ZF-LP scheme can even achieve a higher WSR than the LP-ZF scheme for $L_t = 1.5 \,\mathrm{m}$. It seems that for large $L_t$, the BS MAs can be placed further apart, reducing the correlation between their impulse responses. Regarding the schemes with only fixed antennas, the LP-FIX scheme provides a lower WSR than the LP-MA scheme, and this difference is increased both with $a_k$ and $L_t$. On the other hand, the WSR for the ZF-MA scheme is always negligible.}

\begin{figure}[t]
\centering{}\includegraphics[scale=0.85]{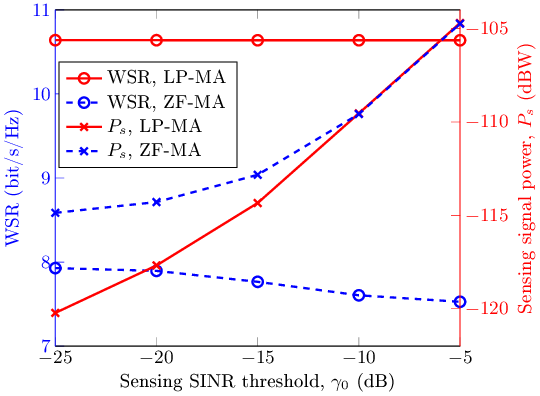}\caption{WSR and sensing signal power versus $\gamma_{0}$.\label{fig:WSR-Ps-gamma0}}
\end{figure}

In Fig. \ref{fig:WSR-Ps-gamma0}, we show the WSR and the sensing
signal power $P_{s}$ versus the sensing SINR threshold $\gamma_{0}$.
In general, we notice that the considered sensing metric is far more
sensitive to the change of the sensing SINR threshold compared the
considered communication metric for both schemes. More precisely,
the WSR for the ZF-MA scheme drops by only 0.4 bits/s/Hz,
while for the LP-MA scheme it remains almost unchanged.
On the other hand, the sensing signal powers for the LP-MA and the ZF-MA schemes are changed by approximately 10~dB and 15~dB, respectively. It is also noteworthy that the ZF-MA scheme achieves a moderately
larger signal sensing power compared to the LP-MA scheme for low $\gamma_{0}$, while for larger $\gamma_{0}$ the signal sensing powers of these two schemes are approximately equal. 
 

\section{Conclusion}
In this paper, we studied the optimization of the WSR in an MA-assisted ISAC system operating in the near field with MAs placed at the BS transmitter and users. To solve this problem, we developed AO-based algorithms that alternately optimize the sensing receive combiner, the communication precoding matrices, the sensing transmit beamformer, the positions of the users’ MAs and the positions of the BS transmit MAs, for the cases of linear and ZF precoding of the communication signal at the BS. The proposed algorithms were also shown to converge in a small number of iterations. Simulation results verified the effectiveness of the proposed schemes in comparison to fixed-antenna benchmark schemes in a near-field ISAC system, particularly for the scheme with ZF precoding. The largest WSR for the scheme with linear precoding is obtained for the unbalanced users’ rate weights, while the WSR for the scheme with ZF precoding {remains stable across} all users’ rate weights. {Furthermore,} we demonstrated that ZF precoding can {outperform linear precoding for a low number of MAs per user, but its WSR can also significantly drop when the user's inter-antenna interference becomes strong.} For a larger number of MAs per user, linear precoding clearly provides a better WSR because of its ability to adaptively allocate power. {Moreover, increasing the sizes of the MA regions can achieve a much greater performance improvement with ZF precoding as compared to schemes with linear precoding.} Finally, it can be observed that the sensing SINR threshold has a significantly greater influence on the sensing than on the communication features of ISAC. Further research may include the evaluation of~the proposed methods for near-field ISAC systems with multiple sensing targets. {Future work will consider the evaluation of the proposed methods under practical limitations of MA-enabled systems,  
{as well as optimization using a shared power budget for communication and sensing functions.}}

\appendices{}

\section*{APPENDIX A: Proof of Lemma \ref{lem:Proof_lem_MA_user}} \label{sec:Grad_qk_Rk}

The communication rate of user $k$ can be rewritten as 
\begin{align}
R_{L,k} & =\ln\left|\mathbf{A}_{1}\right|-\ln\left|\mathbf{A}_{2,k}\right|,
\end{align}
where $\mathbf{A}_{1}=\sum_{u=1}^{K}\mathbf{H}_{k}\mathbf{W}_{u}\mathbf{W}_{u}^{H}\mathbf{H}_{k}^{H}+\mathbf{H}_{k}\mathbf{v}\mathbf{v}^{H}\mathbf{H}_{k}^{H}+\sigma_{k}^{2}\mathbf{I}_{N_{u}}$
and $\mathbf{A}_{2,k}=\mathbf{A}_{1}-\mathbf{H}_{k}\mathbf{W}_{k}\mathbf{W}_{k}^{H}\mathbf{H}_{k}^{H}$.

Differentiating the terms in the previous expressions, we get 
\begin{align}
\text{d}\ln\left|\mathbf{A}_{1}\right|\!= & 2\mathfrak{R}\Big(\sum_{m=1}^{N_{t}}\sum_{b'=1}^{N_{u}}\!\!\mathbf{C}_{1}(m,b')\text{d}\mathbf{H}_{k}('b,m)\Big)\label{eq:dA1},\\
\text{d}\ln\left|\mathbf{A}_{2,k}\right|\!= & 2\mathfrak{R}\!\Big(\sum_{m=1}^{N_{t}}\!\sum_{b'=1}^{N_{u}}\!\!\mathbf{C}_{2,k}(m,b')\text{d}\mathbf{H}_{k}(b',m)\Big)\!.\label{eq:dA2}
\end{align}

Taking the derivatives with respect to individual coordinates of MA
$b$, we have
\begin{align}
\frac{\text{d}\mathbf{H}_{k}(b,m)}{\text{d}x_{k,b}} & =j\frac{2\pi\rho_{k}}{\lambda}\frac{x_{k,b}-x_{m}}{||\mathbf{t}_{m}-\mathbf{q}_{k,b}||}e^{j\frac{2\pi}{\lambda}||\mathbf{t}_{m}-\mathbf{q}_{k,b}||}\label{eq:dHk_x},\\
\frac{\text{d}\mathbf{H}_{k}(b,m)}{\text{d}y_{k,b}} & =j\frac{2\pi\rho_{k}}{\lambda}\frac{y_{k,b}-y_{m}}{||\mathbf{t}_{m}-\mathbf{q}_{k,b}||}e^{j\frac{2\pi}{\lambda}||\mathbf{t}_{m}-\mathbf{q}_{k,b}||}.\label{eq:dHk_y}
\end{align}

Substituting (\ref{eq:dHk_x}) and (\ref{eq:dHk_y}) into (\ref{eq:dA1}) (for $b'=b$),
we obtain
\begin{align}
&\frac{1}{\text{d}x_{k,b}}\text{d}\ln\left|\mathbf{A}_{1}\right|=-\frac{4\pi\rho_{k}}{\lambda}\times\nonumber \\
&\qquad\mathfrak{I}\Big(\sum_{m=1}^{N_{t}}\mathbf{C}_{1}(m,b)\frac{x_{k,b}-x_{m}}{||\mathbf{t}_{m}-\mathbf{q}_{k,b}||}e^{j\frac{2\pi}{\lambda}||\mathbf{t}_{m}-\mathbf{q}_{k,b}||}\Big)
\end{align}
and 
\begin{align}
&\frac{1}{\text{d}y_{k,b}}\text{d}\ln\left|\mathbf{A}_{1}\right|=-\frac{4\pi\rho_{k}}{\lambda}\times\nonumber \\
&\qquad\mathfrak{I}\Big(\sum_{m=1}^{N_{t}}\mathbf{C}_{1}(m,b)\frac{y_{k,b}-y_{m}}{||\mathbf{t}_{m}-\mathbf{q}_{k,b}||}e^{j\frac{2\pi}{\lambda}||\mathbf{t}_{m}-\mathbf{q}_{k,b}||}\Big).
\end{align}

In a similar manner, we have
\begin{align}
&\frac{1}{\text{d}x_{k,b}}\text{d}\ln\left|\mathbf{A}_{2,k}\right|=-\frac{4\pi\rho_{k}}{\lambda}\times\nonumber \\
&\quad\mathfrak{I}\Big(\sum_{m=1}^{N_{t}}\mathbf{C}_{2,k}(m,b)\frac{x_{k,b}-x_{m}}{||\mathbf{t}_{m}-\mathbf{q}_{k,b}||}e^{j\frac{2\pi}{\lambda}||\mathbf{t}_{m}-\mathbf{q}_{k,b}||}\Big),\\
&\frac{1}{\text{d}y_{k,b}}\text{d}\ln\left|\mathbf{A}_{2,k}\right|=-\frac{4\pi\rho_{k}}{\lambda}\times\nonumber \\
&\quad\text{\ensuremath{\mathfrak{I}}}\Big(\sum_{m=1}^{N_{t}}\mathbf{C}_{2,k}(m,b)\frac{y_{k,b}-y_{m}}{||\mathbf{t}_{m}-\mathbf{q}_{k,b}||}e^{j\frac{2\pi}{\lambda}||\mathbf{t}_{m}-\mathbf{q}_{k,b}||}\Big).
\end{align}
Combining the previous expressions, we obtain (\ref{eq:dRk_x}) and
(\ref{eq:dRk_y}).

\section*{APPENDIX B: Proof of Lemma \ref{lem:Proof_lem_MA_TX} \label{sec:Grad_kz_mth_MA}}

Differentiating the expression (\ref{eq:SINR_ref}), we obtain
\begin{align}
\text{d}\kappa_{l} & =\text{d}\text{Tr}(\mathbf{u}^{H}\mathbf{G}\mathbf{M}\mathbf{G}^{H}\mathbf{u}),\nonumber \\
 & =\text{Tr}(\mathbf{M}\mathbf{G}^{H}\mathbf{u}\mathbf{u}^{H}\text{d}\mathbf{G}+\mathbf{u}\mathbf{u}^{H}\mathbf{G}\mathbf{M}\text{d}\mathbf{G}^{H}),\nonumber \\
 & =\rho_{s}\text{Tr}(\mathbf{M}\mathbf{G}^{H}\mathbf{u}\mathbf{u}^{H}\mathbf{f}_{r}\text{d}\mathbf{f}_{t}^{H}+\mathbf{f}_{r}^{H}\mathbf{u}\mathbf{u}^{H}\mathbf{G}\mathbf{M}\text{d}\mathbf{f}_{t}),\nonumber \\
 & =2\rho_{s}\mathfrak{R}\Big(\sum_{i=1}^{N_{t}}\mathbf{e}(i)\text{d}\mathbf{f}_{t}(i)\Big).
\end{align}
For the $m$-th transmit MA, the previous expression is equal to
\begin{align}
\text{d}\kappa_{l} & =2\rho_{s}\mathfrak{R}\left(\mathbf{e}(m)\text{d}\mathbf{f}_{t}(m)\right)\nonumber \\
 & =-\frac{4\pi\rho_{s}}{\lambda}\mathfrak{I}\left(\mathbf{\mathbf{e}}(m)e^{j\frac{2\pi}{\lambda}||\mathbf{t}_{m}-\mathbf{s}_{l}||}\text{d}||\mathbf{t}_{m}-\mathbf{s}_{l}||\right)
\end{align}
and from this we can easily obtain (\ref{eq:dh_xm}) and (\ref{eq:dh_ym}).

\section*{APPENDIX C: Proof of Lemma \ref{lem:WSR_UB}} \label{sec:Proof_WSR_UB}

{The  WSR can be bounded by  
\begin{align}
\text{WSR}\le\sum_{k=1}^{K}w_{k}\ln\Bigg|\mathbf{I}+\frac{1}{\sigma_{k}^{2}}\mathbf{\mathbf{H}}_{k}\mathbf{L}\mathbf{\mathbf{H}}_{k}^{H}\Bigg|+w_{k}\ln\Bigg|\sigma_{k}^{2}\mathbf{I}\Bigg|, 
\label{WSR_UB}
\end{align}
where $\mathbf{L}=\sum_{u=1}^{K}\mathbf{W}_{u}\mathbf{\mathbf{W}}_{u}^{H}+\mathbf{\mathbf{v}}\mathbf{\mathbf{v}}^{H}$ and $\mathrm{Tr}(\mathbf{L})\le P_{\max}+1$.  Defining the \ac{EVD} $\mathbf{\mathbf{H}}_{k}^{H}\mathbf{\mathbf{H}}_{k}=\mathbf{U}_{k}\mathbf{\Lambda}_{k}\mathbf{\mathbf{U}}_{k}^{H},$
where $\mathbf{\Lambda}_{k}=\mathrm{diag}(\lambda_{k,1},\dots,\lambda_{k,N_{u}},0,\dots,0)$,
we obtain
\begin{align}
\ln\Bigg|\mathbf{I}+\frac{1}{\sigma_{k}^{2}}\mathbf{\mathbf{H}}_{k}\mathbf{L}\mathbf{\mathbf{H}}_{k}^{H}\Bigg| & =\ln\Bigg|\mathbf{\mathbf{I}}+\frac{1}{\sigma_{k}^{2}}\mathbf{\Lambda}_{k}^{1/2}\mathbf{U}_{k}\mathbf{L}\mathbf{U}_{k}^{H}\mathbf{\Lambda}_{k}^{1/2}\Bigg|\nonumber \\
&
=\ln\Bigg|\mathbf{\mathbf{I}}+\frac{1}{\sigma_{k}^{2}}\mathbf{\Lambda}_{k}^{1/2}\hat{\mathbf{L}}\mathbf{\Lambda}_{k}^{1/2}\Bigg|, \label{eq:UB_EVD}
\end{align}
where $\hat{\mathbf{L}}_{k}=\mathbf{U}_{k}{\mathbf{L}}\mathbf{U}_{k}^{H}$
and $\mathrm{Tr}(\mathbf{L})=\mathrm{Tr}(\hat{\mathbf{L}}_{k}).$
As for any non-negative definite matrix $\mathbf{A}$, we have $\det(\mathbf{A})\le\prod_{i}\mathbf{A}(i,i)$; thus the expression on the right-hand side of (\ref{eq:UB_EVD}) satisfies the bound
\begin{align}
\ln\Bigg|\mathbf{I}\!+\!\frac{1}{\sigma_{k}^{2}}\mathbf{\Lambda}_{k}^{1/2}\hat{\mathbf{L}}_{k}\mathbf{\Lambda}_{k}^{1/2}\Bigg| & \!\le\!\sum_{i=1}^{N_{u}}\ln\Big(1\!+\!\frac{1}{\sigma_{k}^{2}}\hat{\mathbf{L}}_{k}(i,i)\lambda_{k,i}\Big)\label{DetProdIneq}.
\end{align}
}
{The maximum of the right-hand side in (\ref{DetProdIneq}) is achieved if all the summation
terms are equal. This means $\hat{\mathbf{L}}_{k}(i,i)=(P_{\max}+1)/N_{u}$
and $\lambda_{k,i}=N_{t}|\rho_{k}|^{2}$ for each $i$, since $\sum_{i=1}^{N_{u}}\lambda_{k,i}=||\mathbf{\mathbf{H}}_{k}||^{2}=N_{t}N_{u}|\rho_{k}|^{2}$. Therefore, we have
\begin{equation}
\ln\Bigg|\mathbf{I}+\frac{1}{\sigma_{k}^{2}}\mathbf{\Lambda}_{k}^{1/2}\hat{\mathbf{L}}_{k}\mathbf{\Lambda}_{k}^{1/2}\Bigg|\le N_{u}\ln\left(1+\frac{N_{t}|\rho_{k}|^{2}(P_{\max}+1)}{N_{u}\sigma_{k}^{2}}\right).\label{lndet_ineq_fin}
\end{equation}
Substituting (\ref{lndet_ineq_fin}) into (\ref{WSR_UB}), we obtain (\ref{eq:WSR_UB_expr}).}

\section*{APPENDIX D: Proof of Lemma \ref{lem:dRk_UserMAPos}} \label{sec:dRk_UserMAPos}

Differentiating the achievable rate of user $k$, we obtain
\begin{equation}
\text{d}R_{Z,k}\!=\!\text{Tr}(\mathbf{F}_{k}^{-1})\text{d}\text{\ensuremath{\beta_{e}^{2}}}+\text{Tr}((\mathbf{F}_{k}^{-1}\!-\!\mathbf{E}_{k}^{-1})\text{d}(\mathbf{H}_{k}\mathbf{v}\mathbf{v}^{H}\mathbf{H}_{k}^{H})).\label{eq:dRk}
\end{equation}

For the first term on the right-hand side of (\ref{eq:dRk}), we
have
\begin{align}\label{eq:dRk_first}
\text{Tr}&(\mathbf{F}_{k}^{-1})\text{d}\text{\ensuremath{\beta_{e}^{2}}}=\text{Tr}(\mathbf{F}_{k}^{-1})P_{\max}\text{d}T^{-1}\nonumber, \\
=&P_{\max}T^{-2}\text{Tr}(\mathbf{F}_{k}^{-1})\text{Tr}((\mathbf{H}_{e}\mathbf{H}_{e}^{H})^{-2}\text{d}(\mathbf{H}_{e}\mathbf{H}_{e}^{H})),\nonumber \\
=&P_{\max}T^{-2}\text{Tr}(\mathbf{F}_{k}^{-1})\text{Tr}(\mathbf{B}_{e}\text{d}\mathbf{H}_{e}+\mathbf{B}_{e}^{H}\text{d}\mathbf{H}_{e}^{H}), \\
=&\text{\ensuremath{P_{\max}T^{-2}}Tr}(\mathbf{F}_{k}^{-1})2\mathfrak{R}\Big(\sum_{m=1}^{N_{t}}\sum_{k'=1}^{K}\sum_{b'=1}^{N_{u}}\mathbf{B}_{e}(m,b'_{k})\text{d}\mathbf{H}_{e}(b'_{k},m)\Big),\nonumber
\end{align}
where $b'_{k}=(k'-1)N_{u}+b'$.

For the second term on the right-hand side of (\ref{eq:dRk}), we
have 
\begin{align}
\text{Tr}((\mathbf{F}_{k}^{-1}&-\mathbf{E}_{k}^{-1})\text{d}(\mathbf{H}_{k}\mathbf{v}\mathbf{v}^{H}\mathbf{H}_{k}^{H}))\nonumber \\
=&\text{Tr}((\mathbf{F}_{k}^{-1}-\mathbf{E}_{k}^{-1})(\text{d}\mathbf{H}_{k}\mathbf{v}\mathbf{v}^{H}\mathbf{H}_{k}^{H}+\mathbf{H}_{k}\mathbf{v}\mathbf{v}^{H}\text{d}\mathbf{H}_{k}^{H})),\nonumber \\
=&\text{Tr}(\mathbf{G}_{k}\text{d}\mathbf{H}_{k}+\mathbf{G}_{k}^{H}\text{d}\mathbf{H}_{k}^{H}),\nonumber \\
=&2\mathfrak{R}\Big(\sum_{m=1}^{N_{t}}\sum_{b'=1}^{N_{u}}\mathbf{G}_{k}(m,b')\text{d}\mathbf{H}_{k}(b',m)\Big).\label{eq:dRk_second}
\end{align}

Taking the derivatives with respect to coordinates of the $b$-th
MA of user $k$, we have
\begin{align}
\frac{\text{d}\mathbf{H}_{k}(b,m)}{\text{d}x_{k,b}} & =j\frac{2\pi\rho_{k}}{\lambda}\frac{x_{k,b}-x_{m}}{||\mathbf{t}_{m}-\mathbf{q}_{k,b}||}e^{j\frac{2\pi}{\lambda}||\mathbf{t}_{m}-\mathbf{q}_{k,b}||},\label{eq:dHk_xkb}\\
\frac{\text{d}\mathbf{H}_{k}(b,m)}{\text{d}y_{k,b}} & =j\frac{2\pi\rho_{k}}{\lambda}\frac{y_{k,b}-y_{m}}{||\mathbf{t}_{m}-\mathbf{q}_{k,b}||}e^{j\frac{2\pi}{\lambda}||\mathbf{t}_{m}-\mathbf{q}_{k,b}||}.\label{eq:dHk_ykb}
\end{align}
Substituting (\ref{eq:dHk_xkb}) and (\ref{eq:dHk_ykb}) into (\ref{eq:dRk_first})
and (\ref{eq:dRk_second}) (for $b'=b$), respectively, we obtain
that the gradients of $R_{k}$ w.r.t. the coordinates of the $b$-th
MA of user $k$ are given by (\ref{eq:dRu_xkb}) and (\ref{eq:dRu_ykb}). For the achievable rate of another user $u\neq k$, we have
\begin{equation}
\text{d}R_{u}=\text{Tr}(\mathbf{F}_{u}^{-1}\text{d}\text{\ensuremath{\beta_{e}^{2}}}).
\end{equation}
Repeating the same steps as in (\ref{eq:dRk_first}), we obtain (\ref{eq:dRu_xkb})
and (\ref{eq:dRu_ykb}).

\section*{APPENDIX D: Proof of Lemma \ref{lem:dSINR_UserMAPos}\label{sec:dSINR_UserMAPos}}

Differentiating the expression (\ref{eq:SINR_ref-ZF}), we get
\begin{align}\label{eq:dSINR}
\text{d}\kappa_{z} \!& =\text{d}(\text{\ensuremath{\mathbf{u}}}^{H}\mathbf{G}(\gamma_{0}\mathbf{P}_{e}\mathbf{P}_{e}^{H}-\mathbf{v}\mathbf{v}^{H})\mathbf{G}^{H}\text{\ensuremath{\mathbf{u}}}),\nonumber \\
 & =\gamma_{0}\text{d}(\text{\ensuremath{\beta_{e}^{2}}}\text{\ensuremath{\mathbf{u}}}^{H}\mathbf{G}\mathbf{P}_{1}\mathbf{P}_{1}^{H}\mathbf{G}^{H}\text{\ensuremath{\mathbf{u}}}),\nonumber \\
 & =\gamma_{0}\text{\ensuremath{\mathbf{u}}}^{H}\mathbf{G}\mathbf{P}_{1}\mathbf{P}_{1}^{H}\mathbf{G}^{H}\text{\ensuremath{\mathbf{u}}}\text{d}\text{\ensuremath{\beta_{e}^{2}}} \\
 &\qquad\qquad +\gamma_{0}\text{\ensuremath{\beta_{e}^{2}}}\text{Tr}(\text{\ensuremath{\mathbf{u}}}^{H}\mathbf{G}\text{d}(\mathbf{P}_{1}\mathbf{P}_{1}^{H})\mathbf{G}^{H}\text{\ensuremath{\mathbf{u}}}),\nonumber \\
 & =\gamma_{0}\text{\ensuremath{\mathbf{u}}}^{H}\mathbf{G}\mathbf{P}_{1}\mathbf{P}_{1}^{H}\mathbf{G}^{H}\text{\ensuremath{\mathbf{u}}}\text{d}\text{\ensuremath{\beta_{e}^{2}}}+\gamma_{0}\text{\ensuremath{\beta_{e}^{2}}}2\mathfrak{R}(\text{Tr}(\mathbf{D}\text{d}\mathbf{P}_{1})).\nonumber
\end{align}

Using the derivations from (\ref{eq:dRk_first}), we obtain 
\begin{align}
&\gamma_{0}\text{\ensuremath{\mathbf{u}}}^{H}\mathbf{G}\mathbf{P}_{1}\mathbf{P}_{1}^{H}\mathbf{G}^{H}\text{\ensuremath{\mathbf{u}}}\text{d}\text{\ensuremath{\beta_{e}^{2}}}=\gamma_{0}\text{\ensuremath{\mathbf{u}}}^{H}\mathbf{G}\mathbf{P}_{1}\mathbf{P}_{1}^{H}\mathbf{G}^{H}\text{\ensuremath{\mathbf{u}}}\nonumber \\
&\times2\mathfrak{R}\Big(\sum_{m=1}^{N_{t}}\sum_{k'=1}^{K}\sum_{b'=1}^{N_{u}}\mathbf{B}_{e}(m,b'_{k})\text{d}\mathbf{H}_{e}(b'_{k},m)\Big).
\end{align}
where $b'_{k}=(k'-1)N_{u}+b'$. Substituting (\ref{eq:dHk_xkb}) and
(\ref{eq:dHk_ykb}) into the previous expression (for $k'=k$ and $b'=b$),
we obtain the first terms on the right-hand side of (\ref{eq:dSINR_xkb})
and (\ref{eq:dSINR_ykb}), respectively.

For the second term in (\ref{eq:dSINR}), we have 
\begin{align}
&\gamma_{0}\text{\ensuremath{\beta_{e}^{2}}}2\mathfrak{R} (\text{Tr}(\mathbf{D}\text{d}\mathbf{P}_{1}))\nonumber \\
=&\gamma_{0}\text{\ensuremath{\beta_{e}^{2}}}2\mathfrak{R}(\text{Tr}(\mathbf{D}\text{d}\mathbf{H}_{e}^{H}(\mathbf{H}_{e}\mathbf{H}_{e}^{H})^{-1}\!+\mathbf{D}\mathbf{H}_{e}^{H}\text{d}(\mathbf{H}_{e}\mathbf{H}_{e}^{H})^{-1})),\nonumber \\
=&\gamma_{0}\text{\ensuremath{\beta_{e}^{2}}}2\mathfrak{R}(\text{Tr}((\mathbf{H}_{e}\mathbf{H}_{e}^{H})^{-1}\mathbf{D}\text{d}\mathbf{H}_{e}^{H}-\mathbf{D}\mathbf{H}_{e}^{H}(\mathbf{H}_{e}\mathbf{H}_{e}^{H})^{-2}\text{d}(\mathbf{H}_{e}\mathbf{H}_{e}^{H}))),\nonumber \\
=&\gamma_{0}\text{\ensuremath{\beta_{e}^{2}}}2\mathfrak{R}(\text{Tr}(-\mathbf{H}_{e}^{H}\mathbf{D}\mathbf{H}_{e}^{H}(\mathbf{H}_{e}\mathbf{H}_{e}^{H})^{-2}\text{d}\mathbf{H}_{e}\nonumber \\
&+[(\mathbf{H}_{e}\mathbf{H}_{e}^{H})^{-1}\mathbf{D}-\mathbf{D}\mathbf{H}_{e}^{H}(\mathbf{H}_{e}\mathbf{H}_{e}^{H})^{-2}\mathbf{H}_{e}]\text{d}\mathbf{H}_{e}^{H})),\nonumber \\
=&\gamma_{0}\text{\ensuremath{\beta_{e}^{2}}}2\mathfrak{R}(\text{Tr}(-\mathbf{D}_{1}\text{d}\mathbf{H}_{e}+\mathbf{D}_{2}^{T}\text{d}\mathbf{H}_{e}^{*})),\nonumber \\
=&\gamma_{0}\text{\ensuremath{\beta_{e}^{2}}}\sum_{m=1}^{N_{t}}\sum_{k'=1}^{K}\sum_{b'=1}^{N_{ul}}2\mathfrak{R}(-\mathbf{D}_{1}(m,b'_{k})\text{d}\mathbf{H}_{e}(b'_{k},m)\nonumber \\
&+\mathbf{D}_{2}^{T}(m,b'_{k})\text{d}\mathbf{H}_{e}^{*}(b'_{k},m)).
\end{align}
Substituting (\ref{eq:dHk_xkb}) and
(\ref{eq:dHk_ykb}) into the previous expression (for $k'=k$ and $b'=b$),
we obtain the second terms on the right-hand side of (\ref{eq:dSINR_xkb})
and (\ref{eq:dSINR_ykb}), respectively. This completes the proof.

\vspace{-1.25em}
\balance
\bibliographystyle{IEEEtran}
\bibliography{IEEEabrv,IEEEexample,references}

\end{document}